\documentclass[aos,preprint]{imsart}
\RequirePackage[OT1]{fontenc}

\RequirePackage{amsthm,amsmath,amsfonts,amssymb}
\RequirePackage[numbers,sort&compress]{natbib}
\RequirePackage[colorlinks,citecolor=blue,urlcolor=blue]{hyperref}
\RequirePackage{graphicx}
\RequirePackage{amsmath,amssymb,amsthm,mathtools}
\RequirePackage{graphicx,psfrag,epsf}
\RequirePackage{enumerate}
\RequirePackage{url} 
\RequirePackage{hyperref}%
\RequirePackage{cases}
\RequirePackage{bm,bbm}
\RequirePackage{color}
\RequirePackage{enumerate}
\RequirePackage[shortlabels]{enumitem}
\RequirePackage{lscape}
\RequirePackage{rotating}
\RequirePackage{subcaption}
\RequirePackage{tabularx}
\RequirePackage{multirow}
\RequirePackage{pdflscape}
\RequirePackage{longtable}
\RequirePackage[section]{placeins}
\RequirePackage{fmtcount} 
\RequirePackage{array,multirow}
\RequirePackage{lipsum}
\RequirePackage{tikz}
\RequirePackage{booktabs}
\RequirePackage{changepage}
\RequirePackage{algorithm, algorithmicx, algpseudocode}
\RequirePackage{regexpatch}
\RequirePackage{makecell}%
\RequirePackage[
singlelinecheck=false 
]{caption}
\RequirePackage{bigints}

\makeatletter
\newcommand{\multiline}[1]{%
    \begin{tabularx}{\dimexpr\linewidth-\ALG@thistlm}[t]{@{}X@{}}
        #1
    \end{tabularx}
}
\makeatother

\theoremstyle{plain}

\newtheorem{theorem}{Theorem}[section]

\newtheorem{proposition}{Proposition}[section]
\newtheorem{corollary}{Corollary}[section]
\theoremstyle{definition}

\newtheorem{remark}{Remark}[section]
\newtheorem{assumption}{Assumption}

\def\be {\begin{equation}}
\def\ee{\end{equation}}
\def\bea{\begin{eqnarray}}
\def\eea{\end{eqnarray}}
\def\nn{\nonumber}

 \def\R{{\mathbb R}}

 \def\E{{\mathbb E}}
  
 \def\P{{\mathbb P}}

 \def\bx{\bm{x}}
 \def\bX{\bm{X}}

 \def\bW{\bm{W}}
 \def\bz{\bm{z}}
 \def\bZ{\bm{Z}}

 \def\bu{\bm{u}}
 \def\bv{\bm{v}}
 \def\bV{\bm{V}}

 \def\M{\mathbb{M}}
  
  \def\bbeta{\bm{\beta}}
  \def\bgamma{\bm{\gamma}}
  \def\bdelta{\bm{\delta}}
  \def\btheta{\bm{\theta}}

  \def\ep{\varepsilon}
   
  \def\bsk{\bigskip}

  \def\st{\sum_{t=1}^T}
  
  \def\nn{\nonumber}
  \def\what{\widehat}

  \renewcommand{\sp}[1]{\left( #1 \right)}
  
  \newcommand{\lp}[1]{\left\{  #1 \right\}}

  \newcommand{\norm}[1]{\Vert #1 \Vert}
  \DeclareMathOperator*{\argmin}{\arg\min}

\newcommand{\abs}[1]{\left\lvert#1\right\rvert}

\def\dto{\xrightarrow{d} }
\def\pto{\xrightarrow{p}}

\definecolor{mblue}{rgb}{0.38, 0.51, 0.71} 
\definecolor{darkblue}{RGB}{17, 42, 60} 
\definecolor{mred}{RGB}{175, 49, 39} 

\definecolor{orange}{RGB}{217, 156, 55} 
\definecolor{green}{RGB}{144, 169, 84} 
\definecolor{palegreen}{RGB}{197, 184, 104} 

\definecolor{yellow}{RGB}{250, 199, 100} 
\definecolor{brokenwhite}{RGB}{218, 192, 166} 
\definecolor{brokengrey}{rgb}{0.77, 0.76, 0.82} 

\newcommand{\han}[1]{\textcolor{magenta}{#1}}
\def\t{^{\mathrm{\scriptscriptstyle T}}}
\usepackage{dsfont}
\def\id{\mathds{1}}
\newcommand{\1}[1]{{\mathds{1}}\left\{ #1 \right\}}

\begin{document}

\numberwithin{equation}{section}


\begin{frontmatter}
\title{ Statistical Inference for Four-Regime Segmented Regression Models}
\runtitle{Segmented Regression Model}

\begin{aug}

\author[A]{\fnms{Han}~\snm{Yan}\ead[label=e1]{hanyan@stu.pku.edu.cn}}
\and
\author[B]{\fnms{Song Xi}~\snm{Chen}\ead[label=e2]{{sxchen@tsinghua.edu.cn}}}

\address[A]{Guanghua School of Management,
Peking University\printead[presep={,\ }]{e1}}

\address[B]{
Department of Statistics and Data Science, Tsinghua University\printead[presep={,\ }]{e2}
}
\end{aug}

\begin{abstract}
Segmented regression models offer model  flexibility and interpretability as compared to the global parametric and the nonparametric models, and yet are challenging in both estimation and inference. We consider a four-regime segmented model for temporally dependent data with  segmenting boundaries depending on  multivariate covariates with non-diminishing boundary effects. A mixed integer quadratic programming algorithm is formulated to facilitate the least square estimation of the regression and the boundary parameters. The rates of convergence and the asymptotic distributions of the least square estimators are obtained for the regression and the boundary coefficients, respectively. 
We propose a smoothed regression bootstrap to facilitate inference on the parameters and a model selection procedure to select the most suitable model within the model class with at most four segments.
Numerical simulations and a case study on air pollution in Beijing are conducted to demonstrate the proposed approach, which shows that the segmented models with three or four regimes are suitable for the modeling of the meteorological effects on the PM$_{2.5}$ concentration.

\end{abstract}



\end{frontmatter}

\section{Introduction}
\label{sec:intro}

Regression analysis is  a pivotal tool in modeling the relationship between dependent and independent variables and for prediction purposes. 
It is often conducted via two types of models: the global parametric and local nonparametric models.   
The global parametric models, such as the linear and polynomial regression models,  have the advantages of interpretability and computation simplicity. However, they 
often perform poorly due to model misspecification as the underlying model may change over different parts of the domain. 
To have better adaptability,  nonparametric local models 
facilitated by the kernel smoothing, 
the wavelets or splines,  
or the regression trees,  have been introduced. 
The local model's complexities increase with the data's dimension and the sample sizes, elevating the risk of overfitting. 
The segmented model is a compromise between the global and the local models as they are as interpretable as the global parametric models but have improved model specifications.  

Conventional threshold regression model (also called regime switching model) 
\citep{tong1983threshold} was the first generation of the segmented models. 
It assumes that the regression function is of form $\E(Y | \bX, Z) = \bX\t\bbeta + \bX\t\bdelta \id(Z > r)$, where $Z$ is an observable scalar variable that can be either a time index or a pre-specified
 random variable. 
The threshold regression model has a wide range of applications in empirical research, ranging from 
 modeling effects of shocks to economic systems over the business cycles  \citep{potter1995nonlinear},
the dose-response models in biostatistics 
\citep{schwartz1995threshold}, and in sociological research \citep{card2008tipping}. %
Statistical inference of the threshold regression model with a univariate  splitting variable 
has been 
 well developed. \cite{chan1993}, \cite{hansen1996} and \cite{hansen2000} established asymptotic properties of the least squares estimators of the threshold regression models and proposed tests on the threshold effect. 
  As extensions, \cite{gonzalo2002estimation} and \cite{li-ling-2012} introduced the multiple threshold regression model $\E(Y | \bX,Z) = \bX\t\bbeta+ \sum_{k=1}^{K} \bX\t\bdelta_k \id(Z > r_k)$   with $K$ splits 
   and $K+1$ regimes (segments) and investigated the statistical inference problems. 

A limitation of the existing threshold regression approach is that the splitting threshold is largely determined by a univariate variable  $Z$.  
\cite{auerbach2012} showed difficulties  in finding the univariate splitting variable in the analysis of macroeconomic effects of fiscal policies, and  \cite{jiang2014} indicated  that a univariate $Z$  
was not suitable to regulate the gene effects on disease risks as the risk of developing a particular disease was due to multiple genes.  
Recently, \cite{lee2021} and \cite{yu2021}  extended the threshold regression to allow regime switching  
driven by a  {multivariate} random vector $Z$ which is either observable or obtained via a factor model.
Although these works
overcome the limitation of the univariate split variable, 
the setting of at most two regimes can be restrictive for some applications. 
{Machine learning methods, 
such as the convex piece-wise linear fitting, 
can produce segmented linear regression with {unlimited number of} regimes. However, as these methods were focused on the fitting performances, the underlying segmented models {may}  not be identifiable with the suggested procedures. 
The finite mixture models (FMM) proposed by \cite{khalili2007} can also produce a subgroup linear model fitting for heterogeneous data. However, the subgroups from the FMMs do not lead to parameterized boundaries, and thus are less interpretable than the segmented linear models.}

{Our study is motivated from modelling the meteorological effects on PM$_{2.5}$ concentration in Beijing, 
where a global parametric model is too simple to offer good fitting performances and a nonparametric model may be too local and do not provide sufficient atmospheric interpretation. 
 The air pollution in Beijing is typically governed by different meteorological regimes, namely the removal process by favourable northerly wind which removes PM$_{2.5}$ to a low level, the calm regime between the northerly cleaning and the start of the  transported pollution driven under the southerly wind,  the pollution growth regime under southerly wind that transports  polluted air from the south, and air stagnation regime after the pollution has peaked, followed by the next removal process by the northerly wind. These 
 motivate 
 the four-regime segmented regression model in this work. 
 As the air quality and meteorological data are time series, we consider temporally dependent data in the study. } 

Motivated by the air pollution problem,
we consider 
four-regime regression models whose splitting hypersplines are determined 
by linear combinations of two multivariate covariates $\bZ_1$ and $\bZ_2$, 
where the splitting variables $\bZ_1$ and $\bZ_2$ can be any regressors, and the two splitting hyperplanes can intersect.  These make the four regime-regression model less restrictive than the multiple threshold regression model of \cite{bai1998} and \cite{li-ling-2012} where the splitting variable $Z$ is univariate, {and hence allows not necessarily parallel boundary hyperplanes.}
{The four-regime models  
include two and three regime models as special cases, 
 where the splitting boundaries are either parallel or two adjacent regimes share the same regression coefficient and hence can be merged.

{The main contributions of the study are the following.}
We first establish the consistency and the asymptotic distributions of 
the  least squares estimators (LSEs) for both  the boundary and the regression coefficients under the four-regime regression model with  temporally dependent $\rho$-mixing observations, overcoming challenges posed by (i)  
the irregular objective function,  (ii) the fixed boundary edge effects rather than the diminishing  effects commonly  treated in the literature and (iii) the unconventional form of the asymptotic distribution for the boundary coefficient vector.  
It is found that the  asymptotic distribution of  LSEs  for the boundary coefficients is determined by  the minimizers of a compound multivariate Poisson process, whose jumps depend on the points near the true hyperplanes, and the boundary coefficient estimators 
are asymptotically independent of the regression coefficient estimators.

The generalization to the four regimes with two splitting boundaries brings considerable computational challenges. 
 Although the LSE of the {conventional threshold} regression can be obtained with the grid search method, 
it is not practical in our setting {as the boundaries are defined with multivariate variables}. To overcome the challenges, {we draw inspiration from \cite{bertsimas2016best} and \cite{lee2021} and propose an algorithm based on the mixed integer quadratic programming (MIQP), which is not only computationally efficient but also can be further accelerated by adding an iterative component. It is shown the algorithm can facilitate efficient computation of the LSEs with the rather non-regular form of the {least squares} objective function. } 

{To permit statistical inference, especially in light of the rather unusual asymptotic distribution for the boundary coefficient estimates,  
we develop 
a smoothed regression bootstrap method  
and establish its consistency 
for approximating the distribution of the LSEs.   Furthermore,   
the properties of the LSEs under {degenerated} segmented models with less than four regimes are investigated. In order to find the right segmented models with up to four segments, we  propose a model selection method with a backward elimination procedure 
that is shown to be able to consistently choose the right number of regimes. } 

The paper is organized as follows. Section 2 introduces the four-regime regression model. 
Section 3 presents the theoretical properties and the asymptotic distribution of the LSEs for the regression and boundary parameters. In Section 4, we construct a mixed integer quadratic programming  (MIQP) algorithm to efficiently compute the LSEs. Section 5 considers inference problems for the four-regime regression model. Section 6 investigates the properties of the proposed estimator under degenerated models with less than four regimes {and proposes a model selection method}.  Sections 7 and 8 report simulation and empirical results, respectively. Section 9 conclude the paper with possible extensions. {All technical proofs are relegated to a supplementary material (SM, \cite{yan-chen-2024-SM}). } 

\section{Model setup}
\label{sec:setting}

{We first introduce some notations. 
We use $\mathds{1}(\mathcal{A})$ for the indicator function of an event $\mathcal{A}$,  
$\norm{\bv} = (\sum_{i=1}^d v_i^2)^{1/2}$ for the $L_2$-norm of vector $\bv = (v_1, \cdots, v_d)\t$ and $\mathcal{N}(\bv_0; \delta) = \{\bv: \norm{\bv - \bv_0} \leq \delta \}$ for the $\delta$-neighborhood of $\bv$. 
Define $\bv_{-1}$ as the sub-vector of $\bv$ excluding its first element, i.e., $\bv_{-1} = (v_2, \cdots, v_d)\t$.
 We use $|{E}|$ to denote the cardinality of a set ${E}$. For any two sets ${E}_1$ and ${E}_2$, we denote ${E}_1 \bigtriangleup {E}_2 = ({E}_1 \setminus {E}_2) \cup ({E}_2 \setminus {E}_1)$ as their symmetric difference. 
}

Let $ \{\bW_t = 
(Y_t, \bX_t, \bZ_{1,t}, \bZ_{2,t} )\}_{t=1}^T$ be a sequence of observations, where $Y_t$ is the response variable to covariates $\bX_t \in \R^{p}$ and two partitioning variables  $\bZ_{i, t} \in \R^{d_{i}}$ for $i=1$ and $2$,  which determine the boundaries of the segments or regimes.  {The variables  $\bX_t$,  $\bZ_{1,t}$ and $\bZ_{2,t}$ can share common variables.} The four-regime regression model is  
\begin{equation}
       Y_t =   \sum_{k=1}^4 \bX_t\t\bbeta_{k0} \id\{\bZ_t \in R_k(\bgamma_0)\} + \ep_t, 
    \label{eq:m1}
\end{equation}
where  $\bZ_t$ is the union of variables of $\bZ_{1,t}$ and $\bZ_{2,t}$, $\{\bbeta_{k0}\}_{k = 1}^4$ are the regression coefficients,   $\{\bgamma_{i0}\}_{j=1}^2$ are the boundary coefficients,  $\ep_t$ is the residual satisfying $\E(\ep_t| \bX_t, \bZ_{t}) =0$ {with a finite second moment}, and $ {R}_{k}(\bgamma_{0})$ is the $k$-th region split by the hyperplanes 
$\{H_{i0}: \bz_{i}\t\bgamma_{i0} = 0\}_{i=1}^2$ for $\bz_i \in \R^{d_i}$. The overall parameter of interest is $\btheta = (\bgamma\t,\bbeta\t)\t$ where $\bbeta = (\bbeta_{1}\t, \cdots, \bbeta_{4}\t)\t$ and $\bgamma = (\bgamma_1\t, \bgamma_2\t)\t$. We let $\btheta_0$, $\bbeta_0$ and $\bgamma_0$ denote the respective true parameters. For any observation $\bW_t$, it is the signs of $\bZ_{1,t} \t \bgamma_{10}$ and $\bZ_{2,t} \t \bgamma_{20}$ that determine which regression region it is located at.   Denote by $\id_1(U,V) = \id(U>0, V>0), \id_2(U,V) = \id(U\leq 0, V>0), \id_3(U,V) = \id(U\leq 0, V\leq 0)$ and $\id_4(U,V) = \id(U>0, V \leq 0)$. Then we can write Model \eqref{eq:m1} equivalently as 
\begin{equation}
       Y_t =  \sum_{k=1}^4 \bX_t\t\bbeta_{k0} \id_k(\bZ_{1,t}\t\bgamma_{10}, \bZ_{2,t}\t\bgamma_{20}) + \ep_t,
    \label{eq:m2}
\end{equation}
{which explicitly reflects the role of $\bgamma_{0}$ in Model \eqref{eq:m1}. }

\begin{remark}
\label{remark2-1}
Although the splitting hyperplanes  appears linear, non-linearity 
may be accommodated  
by including nonlinear transformed variables in $\bZ_i (i = 1,2)$, for instance, $\bZ_1 = (Z_1, Z_1^2, 1)\t$. The same extension can be conducted to $\bX$. 
It is also noted that in the special case of {$\bZ_{1,t}$ having the same distribution with $\bZ_{2,t}$}, the four segments under $\bgamma_0 = (\bgamma_{10}\t, \bgamma_{20}\t)\t$ are not distinguishable from that under $\tilde{\bgamma}_0 = (\bgamma_{20}\t, \bgamma_{10}\t)\t$. Consequently, $\btheta_0$ is only identifiable up to some permutations. To avoid such situation,  {we assume that the distributions of $\bZ_{1,t}$ and $\bZ_{2,t}$ are distinct}. 

\end{remark}

\begin{remark}
Since the signs of  
$\bZ_1\t\bgamma_{10}$ and $\bZ_2\t\bgamma_{20}$ determine the regimes in Model (\ref{eq:m1}),  
$\bgamma_{10}$ and $\bgamma_{20}$ have  to be normalized {in order to be identifiable}. For any candidate $\bgamma_{i}$ of $\bgamma_{i0}$, we normalize it by its first element $\gamma_{i,1}$, resulting in $\bgamma_i =: (1, \widetilde{\bgamma_{i}})$ where $\widetilde{\bgamma_{i}}$ is assumed to take values in a compact set. 
As noted in \cite{lee2021}, an alternative normalization is $\norm{\bgamma_i}_2 = 1$. In this study, we employ the former  as it has one less parameter. 
\label{remark: gamma normalization}
\end{remark} 

\section{ Estimation and asymptotic  properties}
\label{sec:properties}

In this section, we  outline the least squares (LS) estimation for $\btheta_0$ 
of the four-regime regression model, and 
establish the convergence rates of the LS estimators for the regression coefficient $\what{\bbeta}$ and the boundary coefficient $\what{\bgamma}$ followed by providing their asymptotic distributions. 

With the data sample  $\{\bW_t = (Y_t, \bX_t, \bZ_{1,t}, \bZ_{2,t})\}_{t=1}^T$, in view of 
$\E(\ep_t | \bX_t, \bZ_t) = 0$, 
we define the following least squares criterion function
\begin{align}
   \mathbb{M}_T(\btheta) =    \frac{1}{T}\sum_{t=1}^T \left\{ Y_t - \sum_{k=1}^4 \bX_t\t \bbeta_k \id_k(\bZ_{1,t}\t\bgamma_{1}, \bZ_{2,t}\t\bgamma_{2})  \right\}^2  =:  \frac{1}{T}\sum_{t=1}^T m(\bW_t, \btheta), 
    \label{eq:ls-1}
\end{align}
and the parameter space is $\Theta = \Gamma_1 \times  \Gamma_2 \times \mathcal{B}^4$, where $\Gamma_i$ is a compact set in $\R^{d_i}$ and  the first element of any $\bgamma \in \Gamma_i$ is normalized as $1$ for each $i = 1,2$, and $ \mathcal{B}$ is a compact set in $\R^p$. 
Since $\M_T(\btheta)$ is strictly convex in $\bbeta$ and piece-wise constant in $\bgamma$ with at most $T$ jumps, it has a unique minimizer $\what{\bbeta} = (\what{\bbeta}_1\t, \cdots, \what{\bbeta}_4\t)\t$ for $\bbeta$, but a set of minimizers for $\bgamma$, which is denoted as $\what{\mathcal{G}}$, such that {a LSE $\what{\btheta} = (\what{\bgamma}\t, \what{\bbeta}\t)\t$ satisfies}  
\begin{align}
    \M_T(\what{\btheta}) =  \inf_{\btheta \in \Theta} \M_T(\btheta)  
    ~~\text{for any}~~\what{\bgamma} \in \what{\mathcal{G}}.
        \label{eq:ls} 
\end{align}
It is noted that for any two $\what{\bgamma}, \what{\bgamma}' \in \what{\mathcal{G}}$, the segmented regimes under the corresponding hyperplanes must be the same, as otherwise the estimated regression coefficients will be distinct. In addition, the set $\what{\mathcal{G}}$ is convex since for each $i = 1$  or $2$,  $\bZ_{i,t}\t\what{\bgamma}_i > 0$ and $\bZ_{i,t}\t\what{\bgamma}_i' > 0$ imply that $\bZ_{i,t}\t\widetilde{\bgamma}_i > 0$
for all  $\widetilde{\bgamma}_i = \alpha\what{\bgamma}_i + (1-\alpha)\what{\bgamma}_i' $ with $\alpha \in [0,1]$. }
{
In the rest of this section, we investigate the properties of the LS estimators $\what{\btheta} = (\what{\bgamma}\t, \what{\bbeta}\t)\t$ with  $\what{\bgamma} \in \what{\mathcal{G}}$.} 

\subsection{Identification and consistency}
Here we discuss the identification of $\btheta_0$ and establish the consistency of the LSEs $\what{\btheta}$.  
{
Let $\bW = (Y, \bX, \bZ_1, \bZ_2)$ follow the stationary distribution $\P_0$ of $\bW_t$, and $q_i = \bZ_i\t\bgamma_{i0}$ for {$i = 1$ and $2$}
to indicate whether $\bZ = (\bZ_1, \bZ_2)$ is located on the true hyperplane $H_{i0}:\bZ_i\t\bgamma_{i0} =0$ or not.
Let $\mathcal{S}(i)$ be the set consisting of index pairs $(k,h)$ if $R_k(\bgamma_0)$ and $R_h(\bgamma_0)$ are two adjacent regions split by $H_{i0}$. Specifically,
$\mathcal{S}(1) = \{(1,2),  (2,1)(3,4), (4,3) \}$ and  $\mathcal{S}(2) =  \{(1,4), (4,1), (2,3), (3,2)\} $} according to the provision in the lines above (2.2). Furthermore, let $\bZ$ be the union vector of variables in $\bZ_1$ and $\bZ_2$

\begin{assumption}[temporal dependence]   (i) The {time series} $\{\bW_t\}_{t \ge 1}$  is strictly stationary and $\rho$-mixing  with the mixing coefficient $\rho(t) \leq c \alpha^t $ for finite positive constants  $c$ and $\alpha \in (0,1)$, where $
  \rho(t) = \sup_{s,t \geq 1} \left\{ \sup\text{Corr}(f,g): f\in \Omega_1^s, g \in \Omega_{s+t}^\infty \right\}$, 
 where  $\Omega_i^j$ denotes the $\sigma$-filed generated by $\{\bW_t: i \leq t \leq j\}$.
   (ii) $\E(\ep_t | \mathcal{F}_{t-1}) = 0$, where $\mathcal{F}_{t-1}$ is a filtration generated by $\{(\bX_i, \bZ_{1i}, \bZ_{2i},\ep_{i-1}): i \leq t\}$.
  \label{assumption: temporal}
\end{assumption}

\begin{assumption}[identification] For $i \in \{1,2\}$ and $k,h \in \{1,\cdots, 4\}$, 
{(i) $\bZ_{1}$ and $\bZ_{2}$ are not identically distributed.  }
(ii)  There exists {a} $j \in \{1, \cdots, d_i\}$ such that $\P(|q_i| \leq \epsilon| \bZ_{-j,i} )   > 0$ almost surely for $\bZ_{-j,i}$ and for any $\epsilon > 0$,  where $\bZ_{-j,i}$ is the  vector after excluding $\bZ_i$'s $j$th element; {without loss of generality, assume $j = 1$. }
(iii) For any $\bgamma \in \Gamma_1 \times \Gamma_2$ and $\P\left\{ \bZ \in {R}_{k}(\bgamma_{0}) \cap {R}_{h}(\bgamma)\right\} > 0$,  the smallest eigenvalue of $\E\left\{\bX\bX\t | \bZ \in {R}_{k}(\bgamma_{0}) \cap {R}_{h}(\bgamma)\right\} \geq \lambda_0$ for some constant $\lambda_0 > 0$. 
(iv)  For $(k, h) \in \mathcal{S}(i)$,  $\norm{\bbeta_{k0} - \bbeta_{h0}}> c_0$ for some constant $c_0 > 0$. 
\label{assumption: indentification}
\end{assumption}

\begin{assumption} 
(i) $\E(Y^4)<\infty$, $\E(\norm{\bX}^4 )< \infty$ and $\max_{i= 1,2}\E(\norm{\bZ_{i}}) < \infty$. 
(ii)
For each {$i  = 1$ and $2$}, ${ \P(\bZ\t_i \bgamma_{1} < 0< \bZ\t_i \bgamma_{2} ) }  \leq c_1 \norm{\bgamma_{1} - \bgamma_{2}}$ if $\bgamma_{1}, \bgamma_{2,} \in \mathcal{N}(\bgamma_{i0}; \delta_0)$, for some constants $\delta_0, c_1 > 0$.  

\label{assumption: for consistency}
\end{assumption}

{Assumption \ref{assumption: temporal} (i) prescribes the strict stationarity and $\rho$-mixing condition on the time series, 
as used in the existing time-series threshold regression literature (\cite{hansen2000} and \cite{lee2021}). 
It is noted that such a decaying rate is only required in deriving the limiting distribution of $\what{\bgamma}$, which can be relaxed to the polynomial decay for Theorem \ref{thm: consistent} and Theorem \ref{thm: rate}. 
Assumption \ref{assumption: temporal} (ii) imposes a martingale difference condition for the noises, which is standard for time series regressions.} 

Assumption \ref{assumption: indentification} is for the identification of $\btheta_0$.  
Specifically, 
{without Assumption \ref{assumption: indentification} (i), $(\bgamma_1\t, \bgamma_2\t)\t$ are not distinguishable from $(\bgamma_2\t, \bgamma_1\t)\t$ as discussed in Remark \ref{remark2-1}.} It is noted that the methods and theories in the rest of the papers are applicable without such a condition, while a permutation for $\bgamma_1$ and $\bgamma_2$ is possibly required. 
Section F of the SM (\cite{yan-chen-2024-SM})  provides sufficient conditions for Assumption \ref{assumption: indentification}  (ii), 
{which ensures there are positive probability of observations located around the true splitting hyperplanes.}
{Discrete variables can be accommodated in $\bZ_i$, as long as it includes at least one continuous variable, say $Z_{1,i}$. Otherwise, if all the splitting variables are discretely distributed, then $\E\{m(\bW, \btheta) \}$ will be piece-wise constant and  $\bgamma_0$ will not be identifiable.}
Assumption \ref{assumption: indentification} 
(iii) guarantees that the splittings by candidate hyperplanes do not lead to degenerated covariance matrices, which is needed for the identification of $\bbeta_0$.
Assumption \ref{assumption: indentification} 
{(iv) means that adjacent regimes have distinguishable regression coefficients so that the splitting effect of each hyperplane is strictly bounded away $0$, which is similar to the fixed threshold effect models treated in \cite{chan1993} and \citep{yu2021}.} {Assumption \ref{assumption: for consistency} (i) is a moment condition, and (ii) means $\P(\bZ_i\t\bgamma < 0)$ is continuous at $\bgamma_{i0}$, implying that $\E\{m(\bW; \btheta) \}$ is continuous at the true parameter $\btheta_0$. } 

\bigskip 

The identification of $\btheta_0$ 
is formally ensured in the following proposition.  


\begin{proposition}
    Under Assumptions \ref{assumption: temporal} and \ref{assumption: indentification}, $\E\{m(\bW, \btheta)\} > \E\{m(\bW, \btheta_0)\}$ for any $\btheta \in \Theta$ and $\btheta \neq \btheta_0$.
    \label{prop: identification}
\end{proposition}

{The proposition {ensures}
that despite the multiple LS estimates $\widehat{\bgamma}$, the underlying $\bgamma_0$ is unique.} {The following theorem shows that 
any LSE estimators $\widehat{\btheta} = (\widehat{\bgamma}\t, \what{\bbeta}\t)\t$ defined in \eqref{eq:ls} are consistent to $\btheta$. It is worth noting that though there exist infinitely many solutions $\widehat{\bgamma}$ which are collected in the convex set $\what{\mathcal{G}}$, the consistency of each $\widehat{\bgamma}$ can be guaranteed, which implies that the solution set $\what{\mathcal{G}}$ is a local neighborhood of $\bgamma_0$ with a shrinking radius.} 

\begin{theorem}
    Under Assumptions \ref{assumption: temporal}--\ref{assumption: for consistency},  
    {let $\what{\btheta} = (\what{\bgamma}\t, \what{\bbeta}\t)\t$ for any $\what{\bgamma} \in \what{\mathcal{G}}$, 
    }, 
    then 
    $\what{\btheta} \pto \btheta_0$ as $T \to \infty$.  
    \label{thm: consistent}
\end{theorem}

With the estimated splitting hyperplanes, each datum can be classified into one of the four estimated regimes $\{{R}_k(\what{\bgamma}) \}_{k=1}^4$. 
Besides the estimation accuracy of 
 $\btheta_0$, the classification accuracy is also an important criterion. 
{
It is shown next that the estimated regime $R_k(\widehat{\bgamma})$ is consistent to the true regime $R_k({\bgamma}_0)$ for each $k = 1, \cdots, 4$. 
}

\begin{corollary}
Under the  conditions of Theorem \ref{thm: consistent},  $\P\{\bZ \in   {{R}}_k(\bgamma_0) \bigtriangleup  {{R}}_k(\widehat{\bgamma})  \} \to 0$ as $T \to \infty$ for all  $k \in \{1, \cdots, 4\}$. 
\label{cor: consistent}
\end{corollary}

\subsection{Convergence rates and asymptotic distributions}

{ We first study the convergence rates of the LSEs 
$\what{\bbeta}$ and $\what{\bgamma}$,  which require 
the following conditions. } 

\begin{assumption}

(i) For  $i  = 1$ and $2$, there exist constants $\delta_1, c_2 > 0$ such that if $\epsilon \in (0, \delta_1)$ then  $\P(|q_{i}| < \epsilon | \bZ_{-1,i}) \geq c_2 \epsilon$ almost surely. 
(ii) For $i  = 1$ and $2$, there exists a neighborhood $\mathcal{N}_i = \mathcal{N}(\bgamma_{i0}; \delta_2)$ {of $\bgamma_{i0}$ }  for some $\delta_2 > 0$, such that $\inf_{\bgamma \in \mathcal{N}_i }\E(\norm{\bX\t\bdelta_{kh,0}} |  \bZ\t_i\bgamma = 0 ) > 0$ almost surely for each $(k,h) \in \mathcal{S}(i)$, where $\bdelta_{kh,0} = \bbeta_{k0} - \bbeta_{h0}$. 
(iii)
$\P(\bZ\t_1\bgamma_1 < 0 < \bZ_1\t\bgamma_2 ,  \bZ\t_2\bgamma_3 < 0 < \bZ_2\t\bgamma_4)  \leq c_3 \norm{\bgamma_1 - \bgamma_2}\norm{\bgamma_3 - \bgamma_4}$ for some constant $c_3 > 0$ if $\bgamma_{1}, \bgamma_2 \in \mathcal{N}_1$ and $\bgamma_{3}, \bgamma_4 \in \mathcal{N}_2$.
(iv)  $\sup_{\bgamma \in \mathcal{N}_i }\E(\norm{\bX}^{8}| \bZ\t_i\bgamma = 0) < \infty $ and $ \sup_{\bgamma \in \mathcal{N}_i }\E(\ep^8 | \bZ\t_i\bgamma = 0) < \infty$ almost surely.

\label{assumption: for rate}
\end{assumption}

Assumption \ref{assumption: for rate} (i) strengthens Assumption \ref{assumption: indentification} (i) and is satisfied when the conditional density  $f_{q_i | \bZ_{-1,i}}(q)$ is continuous and bounded away from $0$ at $q = 0$ almost surely. Assumption \ref{assumption: for rate} (ii) ensures  
there is a jump of the regression surface at the splitting hyperplane, which is similar to Assumption D3 of \cite{yu2021} and Assumption 4.(iii) of \cite{lee2021}. 
   Assumption \ref{assumption: for rate} (iii) controls the probability of data near the cross regions of the two hyperplanes, whose sufficient condition is presented in Section F of the SM (\cite{yan-chen-2024-SM}). 
 Assumption \ref{assumption: for rate} (iv) requires that $\norm{\bX}$ and $\ep$ has a finite moment of the order $8$
 around the hyperplanes. 

The next theorem establishes 
 the rates of convergence of $\what{\bbeta}$ and $\what{\bgamma}$, followed by the convergence rate of the proportions of misclassifications.  

\begin{theorem} 
    Under Assumptions \ref{assumption: temporal}--\ref{assumption: for rate}, 
    $\norm{\what{\bbeta} - \bbeta_0} = O_p(1/\sqrt{T})$  {and} 
    $\norm{\what{\bgamma} - \bgamma_0}= O_p(1/T)$ for any $\what{\bgamma} \in \what{\mathcal{G}}$. 
    \label{thm: rate}
\end{theorem}

\begin{corollary}
Under the conditions of Theorem \ref{thm: rate},  $\P\{\bZ \in   {{R}}_k(\bgamma_0) \bigtriangleup  {{R}}_k(\widehat{\bgamma}) \} = O(1/T)$ for all $k \in \{1, \cdots, 4\}$.
\end{corollary} 

The theorem, whose proof is in Section B of the SM (\cite{yan-chen-2024-SM}), shows that the regression coefficient estimator $\what{\bbeta}$ converges to $\bbeta_0$ at the standard $\sqrt{T}$-rate, while the   
 boundary parameter estimator $\what{\bgamma}$, {despite having multiple solutions,}  converges to $\bgamma_0$  at the faster $T$-rate. The super convergence rate attained by $\what{\bgamma}$ is quite typical  for the boundary parameter estimators, for instance,  
the maximum likelihood estimator for 
the boundary parameter of uniform distributions,  the LS estimator of  
models with a jump in the conditional density  \citep{chernozhukov2004}, 
the threshold regression model \citep{chan1993} and the two-regime regression model with a fixed threshold effect \citep{yu2021}. An intuition for the fast convergence of $\what{\bgamma}$ 
 is that the discontinuity of the regression planes is highly informative for the inference of $\bgamma$.
It is noted that in the {shrinking} 
threshold effect setting $\bbeta_{10}-\bbeta_{20} = \bm{c}T^{-\alpha}$ with $\bm{c} \neq 0$ and $0<\alpha < \frac{1}{2}$ adopted by \cite{hansen2000} and \cite{lee2021}, the  convergence rate of $
\what{\bgamma}$ is slower at $T^{1-2\alpha}$. 

{To present the asymptotic distributions of $\what{\bbeta}$ and $\what{\bgamma}$, we define for each $k \in \{1, \cdots, 4\}$,  
}
\begin{align}
    B_k = \E\lp{\bX\bX\t \id(\bZ \in R_k(\bgamma_0) } ~\text{and}~
\Sigma_k = B_k^{-1}\E\lp{\bX\bX\t\ep^2 \id(\bZ \in R_k(\bgamma_0)} B_k^{-1}. \nn 
\end{align}
Let $q_{i,t} = \bZ_{i,t}\t\bgamma_{i0}$ and $q_i = \bZ_i\t \gamma_{i0}$ for  $i = 1$ and $2$. Denote by    
$s_i^{(k)} = (-1)^{\id(q_{i} \leq 0, ~\forall \bZ \in R_k(\bgamma_0))}$ be the sign of $q_{i}$ for $\bZ = (\bZ_1\t, \bZ_2\t) \in R_k(\bgamma_0)$. For instance, $s_1^{(1)} = s_2^{(1)} = 1$ and $s_1^{(2)} = -1, s_2^{(2)} =1$. If $R_k(\bgamma_0)$ and $R_h(\bgamma_0)$ are adjacent  such that $(k,h) \in \mathcal{S}(i)$ for $i = 1$ or $2$, let    
\bea 
\xi_t^{(k,h)} = \left(\bdelta_{kh,0}\t \bX_t\bX_t\t\bdelta_{kh,0} + 2 \bX_t\t\bdelta_{kh,0}\ep_t \right)\id\left\{\bZ_t \in R_k(\bgamma_0) \cup  R_h(\bgamma_0) \right\}
\eea 
where $\bdelta_{kh, 0} = \bbeta_{k0} - \bbeta_{h0}$. Let $\bZ_{-1,i,t}$ be the random vector of $\bZ_{i,t}$ excluding its first element. 
Suppose $(q_i, \bZ_{-1,i}, \xi^{(k,h)})$ follows the stationary distribution of  $(q_{i,t}, \bZ_{-1,i,t}, \xi^{(k,h)}_t)$. We denote $F_{q_i | \bZ_{-1,i}} (q|\bZ_{-1,i})$ and $F_{\xi^{(k,h)}|q_i,  \bZ_{-1,i}} (\xi|q_i, \bZ_{-1,i})$ as the conditional distributions of $q_i$ on $\bZ_{-1,i}$ and $\xi^{(k,h)}$ on $(q_i,  \bZ_{-1,i})$, respectively, and the corresponding conditional densities are $f_{q_i | \bZ_{-1,i}} (q|\bZ_{-1,i})$ and $f_{\xi^{(k,h)}|q_i,  \bZ_{-1,i}} (\xi|q_i, \bZ_{-1,i})$, respectively. 
{Let $\mathcal{Z}_{-1,i}$ be the support of the distribution of $\bZ_{-1,i}$.}
The following is needed for the weak convergence of $\what{\bgamma}$. 

\begin{assumption} 
{(i) For $i  = 1$ and $2$, there exist constants $\delta_3, c_4 > 0$ such that $\P(|q_{i,t}| \leq \delta_3, |q_{i,t+j}| \leq \delta_3 ) \leq c_4 \left\{ \P(|q_{i,t}| \leq \delta_3 )\right\}^2 $  uniformly for $t \geq 1$ and $j \geq 1$; 
 (ii) For each  $\bz_{-1,i} \in \mathcal{Z}_{-1,i}$, the conditional density  $f_{q_i | \bZ_{-1,i}} (q|\bz_{-1,i})$ is continuous at $q = 0$ and $c_4 \leq f_{q_i | \bZ_{-1,i}} (0|\bz_{-1,i}) \leq c_5$ for some constants $c_4, c_5 > 0$; 
(iii) For each $\xi \in \mathbb{R}$ and $\bz_{-1,i} \in \mathcal{Z}_{-1,i}$, the conditional density $f_{\xi^{(k,h)}|q_i,  \bZ_{-1,i}} (\xi|q_i, \bz_{-1,i})$ is continuous at $q_i = 0$ and $f_{\xi^{(k,h)}|q_i,  \bZ_{-1,i}} (\xi|0, \bz_{-1,i}) \leq c_6$ for a constant $c_6 > 0$; (iv) $\mathcal{Z}_{-1,i}$ is a compact subset of $\R^{d_i-1}$.}
    \label{assumption: for lim dist}
\end{assumption}

{
Assumption \ref{assumption: for lim dist} (i)  is a non-clustering condition that states the probability of two points are both located near the splitting hyperplane $H_{i0}$ is of a smaller order compared to that of just one point is located near $H_{i0}$, which curbs the clustering of extreme events and is similar to  Condition C.4 of \cite{chernozhukov2011inference}. Assumption \ref{assumption: for lim dist} (ii) and (iii) are 
on the conditional densities $f_{q_i | \bZ_{-1,i}}$ and $f_{\xi^{(k,h)}|q_i,  \bZ_{-j,i}}$, respectively, which are used to characterize behaviors of the points near $H_{i0}$.  The compactness of $\mathcal{Z}_{-1,i}$ is required by the limiting theory of point processes (\cite{resnick1987} and \cite{chernozhukov2004}), which may be attained 
by trimming $\bZ_{-1,i,t}$ or 
empirical quantile transformation.
}

\bigskip 
{The asymptotic distribution of $\widehat{\bgamma}$ needs the following stochastic process}
  \begin{align}
{ D(\bv) = \sum_{i = 1, 2}\sum_{{k,h} \in \mathcal{S}(i)} \sum_{\ell=1}^{\infty} \xi_{i,\ell}^{(k,h)} \id\left\{{J}^{(k,h)}_{i, \ell}+ ( \bZ_{i, \ell}^{(k,h)})\t \bv_{-1,i}  \leq 0 <   {J}^{(k,h)}_{i, \ell}\right\}, } 
    \label{eq: D(v)}
  \end{align}
 for $\bv = (\bv_1\t, \bv_2\t)\t \in \R^{d_{1}+d_{2}}$,  where $\{(\xi_{i, \ell}^{(k,h)}, \bZ_{i, \ell}^{(k,h) })\}_{\ell = 1}^\infty$ are independent copies of $(\bar{\xi}^{(k,h)}_i, \bZ_{-1, i})$  with $\bar{\xi}^{(k,h)}_i\sim F_{\xi^{(k,h)}|q_i,  \bZ_{-1,i}} (\xi|0, \bZ_{-1, i})$, {and} ${J}^{(k,h)}_{i, \ell} = \mathcal{J}^{(k,h)}_{i, \ell} / {f_{q_{i}|\bZ_{-1,i}}(0 | \bZ_{i, \ell}^{(k,h) })}$ 
  with $\mathcal{J}^{(k,h)}_{i, \ell} = s_i^{(k)} \sum_{n=1}^{\ell} \mathcal{E}_{i, n}^{(k,h)}$ and $\{ \mathcal{E}_{i, n}^{(k,h)} \}_{n=1}^\infty$ are independent unit exponential variables which are independent of $\{(\xi_{i,\ell}^{(k,h)}, \bZ_{i, \ell}^{(k,h) })\}_{\ell = 1}^\infty$. Moreover, $\{(\xi_{i,\ell}^{(k,h)}, \bZ_{i, \ell}^{(k,h) }, {J}^{(k,h)}_{i, \ell})\}_{\ell = 1}^\infty$ are mutually independent with respect to $i = 1,2$ and $(k,h) \in \mathcal{S}(i)$.

  {Let $\mathcal{G}_D = \{ \bv_m: D(\bv_m) \leq D(\bv)~\text{if}~\bv \neq \bv_m \}$ be the set of minimizers for $D(\bv)$.  Since $D(\bv)$ is a piece-wise constant random function, there are infinitely many elements in $\mathcal{G}_D$. 
  Such a phenomenon also appears in the threshold regression, where the minimizers of the process, that is a special case of \eqref{eq: D(v)}, are attained in an interval, whose left endpoint is commonly used as a representative, which is not applicable to our case since $\mathcal{G}_D$ is a polyhedron. 
  As treated in \cite{yu2021}, we use the centroid of $\mathcal{G}_D$ as the representative. For any set $\mathcal{A}$ of $d$-dimensional vectors,  the centroid of $\mathcal{A}$ is 
  $ C(\mathcal{A}) = \int_{\bv \in \mathcal{A}} \bv d \bv  / {\int_{\bv \in \mathcal{A}} d \bv}$, which can be geometrically interpreted as the center of mass of the set $\mathcal{A}$. 
  Let $\bgamma_D^c = C(\mathcal{G}_D)$ and $\what{\bgamma}^c= C(\what{\mathcal{G}})$, where $\what{\mathcal{G}}$ is the set for LS estimators for $\bgamma$. The former will define the limit of $\what{\bgamma}^c$ as shown in Theorem 3.3.} 
  Numerically, $\what{\bgamma}^c$ can be approximated by the average of $N$ elements of $\what{\mathcal{G}}$ for a sufficiently large $N$.
  The following theorem establishes the asymptotic distributions of $ \sqrt{T}(\what{\bbeta}_k - \bbeta_{k0})$ and $T(\what{\bgamma}^c - \bgamma_0)$.

\begin{theorem}[Asymptotic distribution] 
Under Assumptions \ref{assumption: temporal}-\ref{assumption: for lim dist},
we have 
{(i) 
$\sqrt{T}(\what{\bbeta}_k - \bbeta_{k0})\dto \bm{N}\sp{0, \Sigma_k }$ for $k = 1, \cdots, 4$ and $T(\what{\bgamma}^c - \bgamma_0) \dto  \bgamma^c_D$; 
}
 (ii) $\{\sqrt{T}(\what{\bbeta}_k - \bbeta_{k0})\}_{k=1}^4$ 
 and $\{T(\what{\bgamma}_{i}^c - \bgamma_{i0})\}_{i=1}^2$  are asymptotically independent. 
 \label{thm: asy-dist}
\end{theorem}

\begin{remark}
The limiting process $D(\bv)$ is derived by 
the asymptotics of the point process induced by $\{ (\xi_{t}^{(k,h)}, \bZ_{-1,i,t}, Tq_{i,t})\}_{t=1}^T$.
{The process $D(\bv)$} 
can be regarded as a multivariate compound Poisson process, 
whose jump sizes are $\{\xi_{i,\ell}^{(k,h)}\}_{\ell=1}^{\infty}$ and jump locations are determined by the counting measure induced by $\{(J_{i, \ell}^{(k,h)}, \bZ_{i, \ell}^{(k,h)} )\}_{\ell=1}^{\infty}$. Intuitively, this is because $D(\bv)$ largely relies on those points lying in a local neighborhood of the true splitting hyperplanes, whose $|q_{i,t}|$ are 
{on the order of $O(T^{-1})$,}
which are rare events with their occurrences { asymptotically governed by} a Poisson process. In the case of univariate threshold model where $\bZ_{i} = (Z, 1)\t$ and $\bgamma_{i0} = (1, \gamma_{i0})\t$ so that $\bZ_{-1, i} = 1$ and $q_i = Z-\gamma_{i0}$, it can be  seen that $D(\bv)$ coincides with the compound Poisson process established in \cite{chan1993}. Theorem \ref{thm: asy-dist} also
extends the result of \cite{yu2021} to accommodate the temporal-dependent data and multiple splitting hyperplanes.   
The analysis is technically more involved than the existing literature of the fixed effect threshold regression due to the challenge of the multivariate boundaries and the dependence of the observations. To tackle these challenges, {we exploit large sample theory for the extreme values and point processes (\cite{meyer1973} and \cite{resnick1987}), as well as the epi-convergence in distribution (\cite{knight1999}), which is more  general than the classic uniform convergence in distribution and allows for more general discontinuity, as outlined in the SM (\cite{yan-chen-2024-SM}).} 
{The techniques} used 
in the proof may be used to analyze the asymptotic of other {extreme type} statistics that can be expressed as some functional of a multivariate point process with temporal-dependent sequences. 
\end{remark}

\begin{remark}
\label{remark3.3}
The asymptotic independence of $T(\what{\bgamma}_1^c- \bgamma_{10}  )$ and $T(\what{\bgamma}_2^c - \bgamma_{20} )$ {was} shown for the univaraite multiple-regime threshold model (\citep{li-ling-2012}). Theorem \ref{thm: asy-dist} reveals that this can be extended to {multiple} splitting hyperplanes, provided that the probability of data locating at the crossing region of the two hyperplanes is negligible as reflected in Assumption \ref{assumption: for rate} (iii). As shown in the proof,  the empirical point process induced by $\{ (\xi_t^{(k,h)}, \bZ_{-1,i,t}, Tq_{i,t}), i = 1, 2, (k,h) \in \mathcal{S}(i)\}_{t=1}^T$
is asymptotic Poisson, whose arrivals can be divided into {different} {segments}, 
depending on whether they {belong} to the same pair $(k,h)\in \mathcal{S}(i)$ or not, 
{where $\mathcal{S}(i)$ is the set of index pairs of adjacent regions split by the $i$-th hyperplane. 
Hence,  the limiting Poisson process can be thinned into several asymptotic independent child processes, which further implies the asymptotic independence of $T(\what{\bgamma}_1^c- \bgamma_{10}  )$ and $T(\what{\bgamma}_2^c - \bgamma_{20} )$. As a building block, we established a thinning theorem for Poisson processes for the $\alpha$-mixing sequences, which might be useful in its own right.} 
The asymptotic independence of 
$ \sqrt{T}(\what{\bbeta}_k - \bbeta_{k0})$ and $T(\what{\bgamma}^c - \bgamma_0)$ can be explained by the fact that {the former 
is asymptotically a sum of terms with each term being asymptotically negligible. Hence 
$ \sqrt{T}(\what{\bbeta}_k - \bbeta_{k0})$ should not depend on}  the stochastically bounded number of points near the hyperplanes that determine the distribution of $T(\what{\bgamma}^c - \bgamma_0)$ (\cite{hsing1995asymptotic}). 

It is also noted that the temporal dependence structure of the observed time series does not show up in the asymptotic distributions of $T(\widehat{\bgamma}^c- \bgamma_0)$ and $\sqrt{T}(\widehat{\bbeta}_k - \bbeta_{k0})$. That regarding $\sqrt{T}(\widehat{\bbeta}_k - \bbeta_{k0})$  is due to 
the martingale difference condition $\E(\ep_t | \mathcal{F}_{t-1}) = 0$ as far as the asymptotic variance of $\widehat{\bbeta}_k$ is concerned, which is commonly the case in other related studies \citep{chan1993, li-ling-2012}. 
{That on the $T(\widehat{\bgamma}^c- \bgamma_0)$  is 
because the asymptotic distribution of $\widehat{\bgamma}^c$ is determined by the 
empirical point process induced by the points near the underlying splitting hyperplanes, which satisfies Meyer's condition (\cite{meyer1973}) for rare events of mixing sequences and  ensures the limiting process being Poisson as in the case of independent observations.}

\end{remark}

\section{Computation} 
\label{sec:MIQP}

{The computation of} the LSE for 
$\what{\btheta}$ by minimizing (\ref{eq:ls}) is quite challenging due to  the non-regularity of $m(\bW_t, \btheta)$ that makes the most commonly used optimization algorithms unworkable. We overcome the 
 difficulty via the mixed integer quadratic programming (MIQP), 
{which optimizes a quadratic objective function with linear constraints 
over points in  polyhedral sets  
whose components can be both integer and continuous variables}; see   \cite{bertsimas2005} and \cite{bertsimas2016best} for details. 
For the two-regime regression, \cite{lee2021} expressed the LS problem as an MIQP problem to improve the computation efficiency. The inclusion of the second boundary in the current study brings challenges.  {If formulated directly using the approach}  of \cite{lee2021},  it would make the objective function quartic rather than quadratic.  
We will formulate a MIQP  for the two-boundary problem to facilitate the computation.  

To make the notations compact, we define 
$I_{k,t} = \id\{\bZ_{t} \in R_k(\bgamma)\}$ for any 
candidate $\bgamma = (\bgamma_1\t, \bgamma_2\t)\t$ 
 and $k = 1, \cdots, 4$. Let $X_{t,i}$ be the $i$-th element of $\bX_t$ and $\beta_{k,i}$ be the $i$-th element of $\bbeta_k$.
 {It can be noted that the irregularity of  $\mathbb{M}_T(\btheta)$ in (\ref{eq:ls}) is brought by the indicators $\{I_{k,t}\}$.}
If we define 
$\ell_{k,i,t} = I_{k,t}\beta_{k,i}$ for $i = 1, \cdots, p$,  then $\mathbb{M}_T(\btheta)$ can be expressed as 
\begin{equation}
  \mathbb{V}_T(\bm{\ell}) = \frac{1}{T}\sum_{t=1}^T\sp{Y_t - \sum_{k=1}^4\sum_{i=1}^{p} X_{t,i}\ell_{k,i,t}}^2
\end{equation}
 {which is quadratic with respect to $\bm{\ell} = \{\ell_{k,i,t}: k = 1, \cdots, 4; ~i = 1, \cdots, p; ~t = 1,\cdots, T\}$}. 

{Since the constraints of an MIQP have to be linear,
while $\ell_{k,i,t} = I_{k,t}\beta_{k,i}$ is non-linear, 
it is necessary to introduce 
 linear constraints to ensure that $\{\ell_{k,i,t} \}$ have a one-to-one correspondence  to the 
unknown parameters $\{\bbeta_k\}_{k=1}^4$ and$ \{\bgamma_j\}_{j=1}^2$. }
As $\bbeta_k$ belongs to  a compact set, there exist constants $L_i$ and $U_i$ such that $L_i \leq \beta_{k,i}\leq U_i$. 
By imposing  constraints 
\begin{equation}
  I_{k,t} L_i \leq \ell_{k,i,t} \leq I_{k,t} U_i \quad \mbox{and} \quad  L_i(1-I_{k,t}) \leq \beta_{k,i} - \ell_{k,i,t} \leq U_i(1-I_{k,t}),
   \label{eq:lc1}
\end{equation} 
it can be verified that (\ref{eq:lc1}) holds if and only if $\ell_{k,i,t} = I_{k,t}\beta_{k,i}$ under the condition that $I_{k,t} \in \{0,1\}$.  That $\ell_{k,i,t} = I_{k,t}\beta_{k,i}$ implies (\ref{eq:lc1}) is obvious. {To appreciate the other way,  note that if $I_{k,t} =1$,  $ \ell_{k,i,t} = \beta_{k,i}$; otherwise if $I_{k,t} =0$,  $\ell_{k,i,t} =0$. In either cases, $\ell_{k,i,t} = I_{k,t}\beta_{k,i}$.} 

The next goal is to relate $I_{k,t}=\id\{\bZ_{t} \in R_k(\bgamma)\}$ to the  boundary coefficients $\{\bgamma_j\}_{j=1}^2$. Let $g_{j,t} = \id(\bZ_{j,t}\t\bgamma_j > 0)$. 
{We first express $g_{j,t}$ by linear  constraints in $\bgamma_j$, so as to link $I_{k,t}$ with $g_{j,t}$ via  linear inequalities. }  
 Let $M_{j,t} = \max_{\bgamma\in\Gamma_j}|\bZ_{j,t}\t\bgamma|$ which can be readily computed via linear programming. 
Then, 
\begin{equation}
(g_{j,t} - 1)(M_{j,t} + \epsilon) < \bZ_{j,t}\t\bgamma_j \leq g_{j,t} M_{j,t} 
\label{eq:lc2}
\end{equation}
hold by the definition of $g_{j,t}$, where $\epsilon >0$ is a small predetermined constant. On the other hand, let $g_{j,t}$ be a binary variable that satisfies (\ref{eq:lc2}). Then, $g_{j,t} = 1$ and the first inequality implies that $Z_{j,t}\t\bgamma_k >0$; and $g_{j,t} = 0$ and the second inequality implies that $Z_{j,t}\t\bgamma \leq 0$. Thus, (\ref{eq:lc2}) are equivalent to $g_{j,t} =  \id(\bZ_{j,t}\t\bgamma_j > 0)$.

Finally, 
{we construct constraints which are linear in $\{g_{j,t}\}_{j=1}^2$ and equivalent to $I_{k,t}=\id\{\bZ_{t} \in R_k(\bgamma)\}$.}
Since  each regime $R_k(\bgamma)$ can be written as $R_k(\bgamma) = \{(\bz_1, \bz_2): s_j^{(k)}\bz_j\t\bgamma_j > 0,~j = 1, 2  \}$, where $s_j^{(k)} \in \{-1, 1\}$ is the sign of $\bz_j\t\bgamma_j$ {for} the points belonging in  $R_k(\bgamma)$, we can write $ I_{k,t} = \prod_{j=1}^2\id(s_j^{(k)}\bZ_{j,t}\t\bgamma_j > 0 )$, which can be linked to  $\{g_{j,t}\}_{j=1}^2$ via 
\begin{align}
    I_{k,t} = \prod_{j=1}^2\id\left(s_j^{(k)}\bZ_{j,t}\t\bgamma_j > 0 \right) = \prod_{j=1}^2\left\{ s_{j}^{(k)} g_{j,t} + (1- s_{j}^{(k)}) / 2\right\},
    \label{eq: I-g}
\end{align}
where the first equality is by the definition of $I_{k,t}$, and the second equality can be directly verified. Since the right-hand side of \eqref{eq: I-g} is a product of two factors taking values in $\{0,1\}$, it can be  shown that \eqref{eq: I-g} is equivalent to the following linear constraints
\begin{align}
     { I_{k,t} \geq \sum_{j=1}^2\left\{ s_{j}^{(k)} g_{j,t} + (1- s_{j}^{(k)}) / 2\right\} -1~~\mbox{and}~~
     I_{k,t} \leq s_{j}^{(k)}  g_{j,t} + (1- s_{j}^{(k)}) / 2}. 
     \label{eq:lc3}
\end{align}
for {$j = 1$ and $2$} and $k \in \{1, \cdots, 4\}$. 

In summary, via the linear constraints (\ref{eq:lc1}), (\ref{eq:lc2}) and  (\ref{eq:lc3}), 
 we transform the original LS problem (\ref{eq:m2}) to a MIQP problem formulated 
 as following. 

Let $\bm{g} = \{g_{j,t}: j=1,2, t = 1,\cdots,T\}$,  $\bm{\mathcal{I}} = \{I_{k,t}: k = 1, \cdots, 4, t = 1, \cdots, T \}$ and $\bm{\ell} = \{\ell_{k,i,t}: k = 1,\cdots,4, i = 1,\cdots,p,  t = 1,\cdots,T \}$.  Solve the following problem: 
\begin{equation}
  \min_{\bbeta, \bgamma, \bm{g},\bm{\mathcal{I}},\bm{\ell}} \frac{1}{T}\sum_{t=1}^T\sp{Y_t - \sum_{k=1}^4\sum_{i=1}^{p} X_{t,i}\ell_{k,i,t}}^2 
  \label{eq:miq eq}
\end{equation}

\begin{equation}
\text{subject to } \left\lbrace\begin{aligned}    
  & \bbeta_k \in \mathcal{B}, ~~ \bgamma_j \in \Gamma_j, ~~   
   g_{j,t} \in \{0,1\}, ~~ I_{k,t} \in \{0,1\},~~ L_i \leq \beta_{k,i} \leq U_i, \\
  & (g_{j,t} - 1)(M_{j,t} + \epsilon) < \bZ_{j,t}\t\bgamma_j \leq g_{j,t} M_{j,t}, ~ I_{k,t} L_i \leq \ell_{k,i,t} \leq I_{k,t} U_i, \\ 
  & L_i(1-I_{k,t}) \leq \beta_{k,i} - \ell_{k,i,t} \leq U_i(1-I_{k,t}),\\ 
  & {I_{k,t} \leq s_{j}^{(k)}  g_{j,t} + (1- s_{j}^{(k)}) / 2, ~~  I_{k,t} \geq \sum_{j=1}^2\left\{ s_{j}^{(k)} g_{j,t} + (1- s_{j}^{(k)}) / 2\right\} -1,  }
\end{aligned}\right.
\label{eq: miq_constraints}
\end{equation}
for $k = 1, \cdots, 4, j = 1, 2,  i = 1, \cdots, p$ and $t = 1, \cdots, T.$

{The above optimization problem can be solved quite efficiently with modern mixed integer optimization 
 softwares such as \textsc{Gurobi} and \textsc{Cplex}. 
The next theorem, whose proof is in Section C of the SM (\cite{yan-chen-2024-SM}), shows that the formulated MIQP is equivalent to the original LS problem.

\begin{theorem} 
{ For any small $\epsilon >0$  in (\ref{eq: miq_constraints}), let $\tilde{\btheta} = (\tilde{\bgamma}\t, \tilde{\bbeta}\t)\t$ be a solution of the MIQP defined with \eqref{eq:miq eq} and \eqref{eq: miq_constraints}}, then
  $\mathbb{M}_T(\what{\btheta}) = \mathbb{M}_T(\tilde{\btheta})$ where $\what{\btheta}$ is {a}
  solution in \eqref{eq:ls}.
\label{thm: MIQP}
\end{theorem}

{
Theorem \ref{thm: MIQP} indicates that any $\widetilde{\bgamma}$ satisfying \eqref{eq:miq eq} and \eqref{eq: miq_constraints} is an element of $\what{\mathcal{G}}$, the solution set for the LS estimators for $\bgamma_0$. 
Since for any $\{{g}_{j,t}\} \in \{0,1\}^{2T}$, there are infinitely many $\bgamma_j~(j = 1,2)$ that satisfy the constraint in the second line of \eqref{eq: miq_constraints},  
we can output multiple solutions $\{\widetilde{\bgamma}_n = (\widetilde{\bgamma}_{n1}\t,\widetilde{\bgamma}_{n2}\t)\t \}_{n=1}^N$ of the above MIQP for a sufficiently large $N$, and use their average as an approximation for the centroid $\what{\bgamma}^c$ of the set $\what{\mathcal{G}}$ as advocated in \cite{yu2021}.}  {
We display a scatter plot of the multiple solutions from a simulation experiment reported in Section H.2 of the SM (\cite{yan-chen-2024-SM}), which appeared to be uniformly distributed. However, it requires further investigation to understand the detailed mechanism regarding how the multiple elements of $\what{\mathcal{G}}$ are produced by the MIQP solver.} 
\begin{remark}

{It is noted that the above algorithm requires prior specifications of $(L_i, U_i)$,  the upper and lower bound for $\beta_{k,i}$. In practice, we can first standardize $\{\bX_t\}_{t=1}^T$ and specify a sufficient large parameter interval $(L_i, U_i)$ to ensure it contains the true value. Alternatively, we can employ the data-driven method proposed in \cite{bertsimas2016best} that estimates $\max\{|L_i|, |U_i|\}$ via  the convex quadratic optimization. Besides the proposed MIQP algorithm, the MCMC-based method as used in \cite{yu2021} for the two-regime regression can also be adapted to minimize the LS criterion $\mathbb{M}_T(\btheta)$, which avoids the specification of the parameter bounds but requires more intensive computations since it is a simulation-based method. A comprehensive comparison between the MIQP and MCMC algorithms for segmented regressions would require more work and we leave it to further study. }
\end{remark}

\begin{remark}
    {As indicated in \cite{lee2021}, the MIQP may be slow when the dimension of $\bX_t$ and the sample size $T$ are large. As an alternative, we present a block coordinate descent (BCD) algorithm for the four-regime model in Section C of the SM (\cite{yan-chen-2024-SM}), which minimizes the LS criterion with respect to $\bbeta$ and $\bgamma$ iteratively. At each step, the update for $\bgamma$ given $\bbeta$ is via a mixed integer linear programming (MILP), 
    {which is easier to solve than the MIQP.}
    The update for $\bbeta$ given $\bgamma$ is by linear regression in each candidate regime. Hence, the BCD is computationally more efficient than the MIQP that jointly optimizes $(\bgamma, \bbeta)$. 
    However, there is no guarantee that the BCD  converges to the global optimal solution without a consistent initialization.  Simulations to compare the two algorithms are presented in the SM (\cite{yan-chen-2024-SM}), which show that the BCD with proper initial values can produce close solutions to that of the MIQP with significantly reduced running time.}  
\end{remark}

\section{Smoothed regression bootstrap}
\label{sec: gamma inference}

We now consider the statistical inference problems for $\bbeta_0$ and $\bgamma_0$. The inference for $\bbeta_0$ is quite standard due to the asymptotic normality of $\widehat{\bbeta}$,  while that for the boundary coefficient $\bgamma_0$ is much more challenging since the asymptotic distribution of $T(\widehat{\bgamma}^c - \bgamma_0)$ has a much-involved form and is hard to simulate. 

A natural idea for the inference of $\bgamma_0$ is to employ the bootstrap. However, as shown in \cite{seijo2011} and \cite{yu2014bootstrap}, neither the nonparametric, the residual, nor the wild bootstrap is consistent in approximating the distribution of estimator for the change points in change point models or the threshold in threshold regression models.  
The failure of these bootstrap methods can be explained as follows. As pointed out in Remark \ref{remark3.3}, only the data around the boundary hyperplanes is informative for the inference on $\bgamma_0$. Thus the bootstrap sampling distribution $\widehat{\P}_T$, when conditional on the original data, must approximate the true distribution $\P_0$ in the neighborhood of the true hyperplanes. 
{For the identification of $\bgamma_0$, 
$\P_0$ must have a positive probability on any local region around the underlying boundaries, as reflected in Assumption \ref{assumption: indentification} (ii).
}
However, 
conditional on the original data, the bootstrap distribution $\widehat{\P}_T$ is discrete under either the nonparametric, the residual, or the wild bootstrap, {which fails to mirror $\P_0$}. As a  remedy, 
we present a smoothed regression bootstrap method and prove its theoretical validity.

{Suppose that $Y$} is generated according to the following segmented linear regression model with  heteroscedastic error
\be 
  Y =  \sum_{k=1}^4 \bX\t \bbeta_{0} \id\{\bZ \in R_k(\bgamma_{0})\} + \sigma_0(\bX, \bZ) \, e, \label{eq: model-heter}
\ee
where $e$ {has a continuous distribution and} is independent of $(\bX, \bZ)$ with $\E\left( e\right) = 0$ and $\E\left( e^2\right) = 1$, and $\sigma_0^2(\bX, \bZ)$ is a conditional variance function representing possible heteroskedasticity. {Model \eqref{eq: model-heter} is a refinement of Model \eqref{eq:m1} with more detailed structure  on the residuals.}  {If it is believed that the error is homogeneous within each region $R_k(\bgamma_0)$ so that $\ep = \sigma_k \id\{\bZ \in R_k(\bgamma_{0})\}e$ for some $\sigma_k > 0$, as assumed in \cite{yu2014bootstrap}, then the nonparametric estimation for $\sigma_0(\bx, \bz)$ is not required and $\sigma_k$ can be estimated with the sample standard deviation of the fitted residuals in the $k$-th region.}

Let $F_0(\bx, \bz)$ be the distribution function of $(\bX, \bZ)$, whose density function is $f_0(\bx, \bz)$. } 
We estimate $F_0(\bx, \bz)$ and $\sigma_0(\bx, \bz)$ nonparametrically with the kernel smoothing. Specifically, let $K_1(\cdot)$ and $K_2(\cdot)$ be a $p$-dimensional and a $(d_1+ d_2)$-dimensional kernel functions, respectively. 
Let $G_i(\bu) = \int_{-\infty}^{\bu} K_i(\bu) d\bu$ for $i = 1, 2$.
{The kernel smoothing estimator for $F_0(\bx, \bz)$ 
is  given by
  \[
\widetilde{F}_0(\bx, \bz) = 
\frac{1}{T} \sum_{t=1}^T  G_1\left(\frac{\bX_t - \bx}{h_1} \right) G_2\left(\frac{\bZ_t - \bz}{h_2} \right),
\]
where $h_1$ and $h_2$ are smoothing  bandwidths.}

With the LS estimator $(\what{\bgamma}, \what{\bbeta})$, the estimated residuals 
are $\what{\ep}_t = Y_t - \sum_{k=1}^4 \bX\t \widehat{\bbeta} \id\{\bZ_t \in R_k(\widehat{\bgamma})\}$. The conditional variance function $\sigma_0^2(\bx, \bz)$ can be estimated via the local linear approach proposed by \cite{fan-yao-1998}. For any given  $(\bx, \bz)$, the local linear estimator $\widetilde{\sigma}^2(\bx, \bz) = \what{\alpha}$, which is defined by 
\[
(\widehat{\alpha}, \widehat{\bm{\eta}})  = \argmin_{(\alpha, \bm{\eta})} \sum_{t=1}^T \left\{\what{\ep}_t^2 -  \alpha - \left((\bX_t - \bx)\t, (\bZ_t - \bz)\t\right) \bm{\eta}  \right\}^2 K_1\left(\frac{\bX_t - \bx}{b_1} \right) K_2\left(\frac{\bZ_t - \bz}{b_2} \right), 
\]
where  $\bm{\eta} \in \R^{p + d_1 + d_2}$, and $b_1$ and $b_2$ are smoothing bandwidths. 
Let $\widehat{e}_{t} = \widehat{\varepsilon}_{t} / \widetilde{\sigma}(\bX_{t}, \bZ_{t})$ and $\tilde{e}_{t} =  \widehat{e}_{t} -  \bar{e}_{T}$, where $\bar{{e}}_T = \sum_{t=1}^T \widehat{e}_{t} / T$.  Denote $\what{{G}}(e)$ as the empirical distribution of  $\{ \tilde{e}_{t} \}_{t=1}^T$.  

\bsk 

{We need the following conditions on the underlying stationary distribution and its density functions, the kernel functions, and the smoothing bandwidths to facilitate the Bootstrap procedure. } 
\begin{assumption} 

(i) The stationary distribution $F_0$ of $(\bX_t,\bZ_t)$ has a compact support and is absolute continuous with density  $f_0(\bx, \bz)$ which is bounded  and  
$ \inf_{\bx, \bz} f_0(\bx, \bz) > 0 $. 

(ii) The conditional variance function $\sigma_0^2(\bx, \bz)$ is bounded and  $ \inf_{\bx, \bz} \sigma_0^2(\bx, \bz) > 0$. 

(iii) The kernels $K_1(\cdot)$ and $K_2(\cdot)$ are symmetric density functions which are Lipshitz continuous and have bounded supports. {The  smoothing bandwidths satisfy  $h_i, b_i \to 0$ for $i = 1$ and $2$, and $T (\log T)^{-1} h_{1}^{p}h_2^{d_1+d_2} \to \infty$ and $T (\log T)^{-1} b_{1}^{p}b_2^{d_1+d_2} \to \infty$ as $T \to \infty$.}

\label{assumption: smooth bootstrap}
\end{assumption}

Under Assumptions \ref{assumption: temporal} and \ref{assumption: smooth bootstrap}, {it can be shown that $\sup_{\bx, \bz}\norm{\widetilde{F}_0(\bx, \bz) - F_0(\bx, \bz)} \pto 0$, 
and $\sup_{\bx, \bz}\norm{\widetilde{\sigma}^2(\bx, \bz) - {\sigma}_0^2(\bx, \bz)} \pto 0$, 
following the uniform convergence results of kernel density and regression estimators for mixing sequences, say  \cite{gyorfi1989}.
{In addition, the above assumptions also ensure the uniform convergence of the density $\widetilde{f}_0$ of the kernel estimator $\widetilde{F}_0$ to the true density function $f_0$, which is required in establishing the consistency of the smoothed regression bootstrap.}
 If $(\bX, \bZ)$ is of high dimensions we can also employ machine learning methods that are adaptive to high dimensional features, such as 
 the deep neural networks, to estimate $f_0(\bx, \bz)$ and $\sigma_0(\bx, \bz)$, as long as their uniform convergence can be guaranteed.

The bootstrap procedure to approximate the distributions of $\{T(\widehat{\bgamma}^c - \bgamma_0), \sqrt{T}(\widehat{\bbeta} - \bbeta_0)\}$ is as follows. 

\vspace{0.3cm}
\textit{Step 1}:  First, generate  $\{(\bX_{t}^*, \bZ_{t}^*)\}_{t=1}^T$ independently from $\widetilde{F}(\bx, \bz)$ and $\{ e_{t}^*\}_{t=1}^T$ independently from $\what{{G}}(e)$, respectively. Then, generate $
Y_{t}^* =    \sum_{k=1}^4 \left(\bX_{t}^*\right)\t \widehat{\bbeta}_k \id\{\bZ_{t}^* \in R_k(\widehat{\bgamma}^c)\} + \widetilde{\sigma}(\bX_{t}^*, \bZ_{t}^*) e_{t}^*$ to obtain bootstrap resample $\{(Y_{t}^*,\bX_{t}^*, \bZ_{t}^*)\}_{t=1}^T$. 

\textit{Step 2}: Compute the LSEs 
based on $\{(Y_{t}^*,\bX_{t}^*, \bZ_{t}^*)\}_{t=1}^T$,  {where $\what{\bbeta}^*$ is the LSE for $\bbeta_0$ and $\{\what{\bgamma}^{*}_i \}_{i=1}^N$ are the LSEs for $\bgamma_0$ for a sufficiently large $N$. Let $\what{\bgamma}^{*c} = \sum_{i=1}^N \what{\bgamma}^{*}_i / N$.}

\textit{Step 3}: Repeat 
the above two steps
$B$ times  for a large positive integer $B$ to obtain $\{\widehat{\bgamma}^{*c}_b\}_{b=1}^B$ and $\{\widehat{\bbeta}^*_b \}_{b=1}^B$, and use the empirical distribution of $\left\{ T (\widehat{\bgamma}^{*c}_b -\widehat{\bgamma}^c), \sqrt{T} (\widehat{\bbeta}^*_b - \what{\bbeta})\right\}_{b=1}^B$ as an estimate of the distribution of $\{T\left(\widehat{\bgamma}^c- \bgamma_0\right), \sqrt{T}(\widehat{\bbeta} - \bbeta_0) \}$. 
\vspace{0.3cm}



{
As in the original LS problem, the LSEs for $\bgamma_0$ based on each bootstrap resample are attained on a convex set $\what{\mathcal{G}}^*$. Therefore, in Step 2 we approximate the centroid of $\what{\mathcal{G}}^*$ by the average of $N$ elements in $\what{\mathcal{G}}^*$.  
Denote the distribution of  $\{T\left(\widehat{\bgamma}^c- \bgamma_0\right), \sqrt{T}(\widehat{\bbeta} - \bbeta_0) \}$ as $\mathcal{L}_T$ and the empirical distribution of $\left\{ T (\widehat{\bgamma}^{*c}_b -\widehat{\bgamma}^c), \sqrt{T} (\widehat{\bbeta}^*_b - \what{\bbeta})\right\}_{b=1}^B$ as  $\mathcal{L}_{T,B}$.} The validity of the smoothed regression bootstrap is established in the following theorem.  
\begin{theorem}
\label{thm: smooth bootstrap}
    Suppose that Assumptions \ref{assumption: temporal}-\ref{assumption: smooth bootstrap} hold. 
    Then $\rho\left(\mathcal{L}_{T,B}, \mathcal{L}_T\right) \pto 0$ as $B, T \to \infty$, for any metric $\rho$ that metrizes weak convergence of distributions. 
\end{theorem}
The proof of the theorem is in Section D of the SM (\cite{yan-chen-2024-SM}) by first establishing sufficient conditions for a consistent bootstrap scheme for approximating  $\mathcal{L}_T$, 
followed by showing that the smoothed regression bootstrap satisfies these conditions. 
With the above result, confidence regions and hypothesis testings about $\bgamma_0$ and $\bbeta_0$ can be readily conducted via the empirical distribution of the smoothed bootstrap estimates $\mathcal{L}_{T,B}$.   

\begin{remark} We exploit the parametric regression model in the bootstrap resampling, under which the mixing-dependent structure of the observed data does not show up in the asymptotic distributions as shown in Theorem \ref{thm: asy-dist}. 
As discussed in 
\cite{hardle2003}, if one has a parametric model that reduces the data generating process to independence sampling, then the parametric bootstrap has properties that are essentially the same as they are when the observations are independently distributed. 
Therefore, in the resampling procedure, the temporal dependence of the original data is not necessary to be explicitly taken into account. 
\end{remark}

\begin{remark}
    {In addition to the smoothed regression bootstrap,  there are two alternative methods which may be applicable for inference of $\bgamma_0$. 
    One is the block subsampling method proposed by   \cite{politis1994-aos},  which  
    was adopted by \cite{gonzalo2005subsampling} 
    in the threshold autoregressive models.  
    Another is the nonparametric posterior confident interval approach based on the Markov Chain Monte Carlo (MCMC)  adopted by \cite{yu2021} for inference on the two-regime regression model. 
    Whether these methods work for the current four-regime segmented regression with fixed boundary effects and dependent data are interesting future research topics. 
    }
\end{remark}

\section{Degenerated models and model selection}
\label{sec: Model Specification}

Model (\ref{eq:m1}) assumes that there are four segments divided by two boundary hyperplanes where the adjacent regimes have distinct regression coefficients. However, it is possible that the underlying regimes are degenerated with less than four regimes.  In this section, we show that the  LS estimator (\ref{eq:ls}) {attains desirable convergence}  
properties even in the degenerated cases, 
{and propose {a} model 
selection method for choosing the underlying model. }  

Given the data sample  $\{(Y_t, \bX_t, \bZ_{1,t}, \bZ_{2,t})\}_{t=1}^T$ for $\bZ_{1,t} \in \mathcal{Z}_{1}$ and $\bZ_{2,t} \in \mathcal{Z}_{2}$,   {there are five possible degenerated models as follows in addition to the four regime model (\ref{eq:m1}).}  

 {\bf (a.1).} Three-regime model with non-intersected splitting hyperplanes:
\begin{equation}
    Y_t = \sum_{k=1}^3 \bX_t\t\bbeta_{k0} \1{\bZ_{t} \in R_k(\bgamma_{0})}+ \ep_t, 
    \label{eq:3-regimes-1}
\end{equation}
where the two  hyperplanes $H_1$ and $H_2$ have no intersection on $\mathcal{Z}_{1}\times \mathcal{Z}_2$. 
Without loss of generality, we  suppose that $\bz_{1}\t\bgamma_{10} \leq  \bz_{2}\t\bgamma_{20}$ for all $(\bz_{1}, \bz_2) \in (\mathcal{Z}_{1} \times \mathcal{Z}_2)$. Then, 
$R_1(\bgamma_{0})= \{\bz: \bz_{1}\t\bgamma_{10} > 0\}, R_2(\bgamma_{0}) = \{\bz: \bz_{1}\t\bgamma_{10} \leq 0, \bz_{2}\t\bgamma_{20} >0 \}$ and $R_3(\bgamma_{0}) = \{\bz: \bz_{2}\t\bgamma_{20}\leq 0 \}$. The conventional multi-threshold models (e.g., \cite{gonzalo2002estimation} and \cite{li-ling-2012}) correspond to this case. 

 {\bf (a.2).} Three-regime regression model with intersected splitting hyperplanes: 
\begin{equation}
    Y_t = \sum_{k=1}^3 \bX_t\t\bbeta_{k0} \id{(\bZ_t \in R_k(\bgamma_{0})}+ \ep_t,  
    \label{eq:3-regimes-2}
\end{equation}
where $R_1(\bgamma_{0})= \{\bz: \bz_{i}\t\bgamma_{j,0} > 0, \bz_{j}\t\bgamma_{j,0} > 0 \}, R_2(\bgamma_0) = \{\bz: \bz_{i}\t\bgamma_{j,0} > 0, \bz_{j}\t\bgamma_{j,0} \leq 0  \}$ and $ R_3(\bgamma_0) = \{\bz: \bz_{j}\t\bgamma_{j,0} \leq 0 \}$ for $i\neq j \in \{1,2\}$.  
{Geometrically, one side of the hyperplane $H_j: \bz_j\t\bgamma_{j,0} = 0$ is split by  $H_i: \bz_i\t\bgamma_{i,0} = 0$ that does not extend to the other side of $H_j$.}

 {\bf (b.1).} Two-regime regression model with one splitting hyperplane:
\begin{equation}
    Y_t = \sum_{k=1}^2 \bX\t_t\bbeta_{k0} \1{\bZ_t \in R_k(\bgamma_{0})}+ \ep_t,  
    \label{eq:2-regimes-1}
\end{equation}
where $(\bz, \bgamma_{0})$ is either $(\bz_1, \bgamma_{10})$ or  $(\bz_2, \bgamma_{20})$ and  $R_1(\bgamma_0) = \{ \bz: \bz\t\bgamma_{0} > 0\}$ and $R_2(\bgamma_0) = \{ \bz: \bz\t\bgamma_{0} \leq 0\}$, which are the same as the two-regime  models of \cite{lee2021} and \cite{yu2021}.

{{\bf (b.2).} Two-regime regression model with two splitting hyperplanes:
\begin{equation}
    Y_t = \sum_{k=1}^2 \bX\t_t\bbeta_{k0} \1{\bZ_t \in R_k(\bgamma_{0})}+ \ep_t,  
    \label{eq:2-regimes-2}
\end{equation}
where $R_1(\bgamma_0) = \{\bz: \bz_1\t\bgamma_{10} > 0, \bz_2\t \bgamma_{20} > 0 \}$ and $R_2\{\bgamma_0 \} = \mathcal{Z}_1 \times \mathcal{Z}_2 \setminus R_1(\bgamma_0)$.} 

 {\bf (c).} Global linear model: 
\begin{equation}
 Y_t =\bX\t_t\bbeta_{0} + \ep_t,
    \label{eq:1-regime}
\end{equation}

\begin{figure}[htbp]
  \centering
  \captionsetup[subfigure]{labelformat=empty}
\begin{subfigure}{0.23\textwidth}
  \caption{ (A): four-regime }
 \centering
    \begin{tikzpicture}
\draw (1.5, 1.5)coordinate(A)-- (-1.5, 1.5)coordinate(B) -- (-1.5, -1.5)coordinate(C) -- (1.5, -1.5)coordinate(D) --cycle;
  \draw[name = H2, color = red] (0,-1.5)coordinate(E) -- (0, 1.5)coordinate(F);
  \draw[name = H1, color = mblue] (-1.5, 0)coordinate(G) --(1.5,-0)coordinate(H) ;  
\coordinate (I) at (intersection of E--F and G--H);
\fill[mblue!30](A)--(F)--(I)--(H)--cycle;
\fill[mred!30](B)--(F)--(I)--(G)--cycle;
\fill[yellow!30](C)--(E)--(I)--(G)--cycle;
\fill[green!30](D)--(E)--(I)--(H)--cycle;
  \node at (-0.33, 1.3) {\textcolor{mred}{\small $H_{1}$}};
  \node at (1.2, 0.2) {\textcolor{mblue}{\small $H_{2}$}};
  \draw[name = H2, color = mred, line width = 0.3mm] (0,-1.5)coordinate(E) -- (0, 1.5)coordinate(F);
  \draw[name = H1, color = mblue, line width = 0.3mm] (-1.5, 0)coordinate(G) --(1.5,-0)coordinate(H) ;  
    \node at (0.75, 0.75){{\small$\underset{(+,+)}{R_1}$}};
    \node at (-0.75, 0.75){{\small$\underset{(-,+)}{R_2}$}};
    \node at (-0.75, -0.75) {{\small$\underset{(-,-)}{R_3}$}};
    \node at (0.75, -0.75) {{\small$\underset{(+,-)}{R_4}$}};
\end{tikzpicture}
\end{subfigure}
\begin{subfigure}{0.23\textwidth}
 \centering
 \caption{ (B): three-regime (a.1) }
    \begin{tikzpicture}
\draw (1.5, 1.5)coordinate(A)-- (-1.5, 1.5)coordinate(B) -- (-1.5, -1.5)coordinate(C) -- (1.5, -1.5)coordinate(D) --cycle;
  \draw[name = H2, color = mred] (-0.5,-1.5)coordinate(E) -- (-0.5,1.5)coordinate(F);
  \draw[name = H1, color = mblue] (0.5, -1.5)coordinate(G) --(0.5, 1.5)coordinate(H) ;  
  \fill[mblue!30](A)--(H)--(G)--(D)--cycle;
\fill[mred!30](H)--(G)--(E)--(F)--cycle;
\fill[yellow!30](E)--(F)--(B)--(C)--cycle;
  \draw[name = H1, color = mblue, line width = 0.3mm]  (-0.5,-1.5)coordinate(E) -- (-0.5,1.5)coordinate(F);
  \draw[name = H2, color = mred, line width = 0.3mm] (0.5, -1.5)coordinate(G) --(0.5, 1.5)coordinate(H) ; 
  \node at (-0.83, 1.3) {\textcolor{mblue}{{\small $H_{2}$}}};
  \node at (0.17, 1.3) {\textcolor{mred}{{\small $H_{1}$}}};
  \node at (1, 0){\makecell[c]{\small $\underset{(+,+)}{R_1}$}};
    \node at (0, 0){{\small$\underset{(-,+)}{R_2}$}};
    \node at (-1, 0) {{\small$\underset{(-,-)}{R_3}$}};
\end{tikzpicture}
\end{subfigure}
\begin{subfigure}{0.23\textwidth}
  \caption{ (C): three-regime (a.2) }
 \centering
    \begin{tikzpicture}
\draw (1.5, 1.5)coordinate(A)-- (-1.5, 1.5)coordinate(B) -- (-1.5, -1.5)coordinate(C) -- (1.5, -1.5)coordinate(D) --cycle;
  \draw[name = H2, color = mred] (0,0)coordinate(E) -- (0, 1.5)coordinate(F);
  \draw[name = H1, color = mblue] (-1.5, 0)coordinate(G) --(1.5,-0)coordinate(H) ;  
\fill[mblue!30](A)--(F)--(E)--(H)--cycle;
\fill[mred!30](B)--(F)--(E)--(G)--cycle;
\fill[yellow!30](C)--(G)--(H)--(D)--cycle;
  \node at (-0.33, 1.3) {\textcolor{mred}{{\small $H_{1}$}}};
  \node at (1.2, 0.2) {\textcolor{mblue}{{\small $H_{2}$}}};
  \draw[name = H2, color = mred, line width = 0.3mm] (0,0)coordinate(E) -- (0, 1.5)coordinate(F);
  \draw[name = H1, color = mblue,line width = 0.3mm] (-1.5, 0)coordinate(G) --(1.5,-0)coordinate(H) ;  
    \node at (0.75, 0.75){{\small$\underset{(+,+)}{R_1}$}};
    \node at (-0.75, 0.75){{\small$\underset{(-,+)}{R_2}$}};
    \node at (0, -0.75) {{\small$\underset{(-,-)\cup(+,-) }{R_3}$}};
\end{tikzpicture}
\end{subfigure}


\begin{subfigure}{0.23\textwidth}
 \centering
 \caption{ (D): two-regime (b.1) }
    \begin{tikzpicture}
\draw (1.5, 1.5)coordinate(A)-- (-1.5, 1.5)coordinate(B) -- (-1.5, -1.5)coordinate(C) -- (1.5, -1.5)coordinate(D) --cycle;
  \draw[name = H1, color = mred] (0,-1.5)coordinate(E) -- (0,1.5)coordinate(F);
  \fill[mblue!30](A)--(F)--(E)--(D)--cycle;
\fill[mred!30](E)--(F)--(B)--(C)--cycle;
  \draw[name = H1, color = mred, line width = 0.3mm] (0, -1.5)coordinate(E) --(0, 1.5)coordinate(F) ; 
  \node at (-0.33, 1.3) {\textcolor{mred}{{\small $H_{1}$}}};
  \node at (0.75, 0){\makecell[c]{\small $\underset{(+)}{R_1}$}};
    \node at (-0.75, 0) {{\small$\underset{(-)}{R_2}$}};
\end{tikzpicture}
\end{subfigure}
\begin{subfigure}{0.23\textwidth}
  \caption{ (E): two-regime (b.2) }
 \centering
    \begin{tikzpicture}
\draw (1.5, 1.5)coordinate(A)-- (-1.5, 1.5)coordinate(B) -- (-1.5, -1.5)coordinate(C) -- (1.5, -1.5)coordinate(D) --cycle;
  \draw[name = H2, color = red] (0,-1.5)coordinate(E) -- (0, 1.5)coordinate(F);
  \draw[name = H1, color = mblue] (-1.5, 0)coordinate(G) --(1.5,-0)coordinate(H) ;  
\coordinate (I) at (intersection of E--F and G--H);
\fill[mblue!30](A)--(F)--(I)--(H)--cycle;
\fill[mred!30](H)--(I)--(F)--(B)--(C)--(D)--cycle;

  \node at (-0.33, 1.3) {\textcolor{mred}{\small $H_{1}$}};
  \node at (1.2, 0.2) {\textcolor{mblue}{\small $H_{2}$}};
  \draw[name = H2, color = mred, line width = 0.3mm] (0,0)coordinate(I) -- (0, 1.5)coordinate(F);
  \draw[name = H1, color = mblue, line width = 0.3mm] (0, 0)coordinate(I) --(1.5,-0)coordinate(H) ;  
    \node at (0.75, 0.75){{\small$\underset{(+,+)}{R_1}$}};
    \node at (-0, -0.5){{\small$\underset{(-,+) \cup (-,-) \cup (+,-)}{R_2}$}};
\end{tikzpicture}
\end{subfigure}
\begin{subfigure}{0.23\textwidth}
  \caption{ (F): global model }
 \centering
    \begin{tikzpicture}
\draw (1.5, 1.5)coordinate(A)-- (-1.5, 1.5)coordinate(B) -- (-1.5, -1.5)coordinate(C) -- (1.5, -1.5)coordinate(D) --cycle;
  \draw[name = H2, color = red] (0,-1.5)coordinate(E) -- (0, 1.5)coordinate(F);
  \draw[name = H1, color = mblue] (-1.5, 0)coordinate(G) --(1.5,-0)coordinate(H) ;  
\coordinate (I) at (intersection of E--F and G--H);
\fill[mblue!30](A)--(B)--(C)--(D)--cycle;
\node at (0, 0){{\small${R_1}$}};
\end{tikzpicture}
\end{subfigure}
\caption{Illustrations of segmented models with no more than four regimes. The signs of $(\bz\t_1\bgamma_{1}, \bz\t_2\bgamma_{2})$ for each region are indicated below the region names.  }
\label{fig: degenerate}
\end{figure}
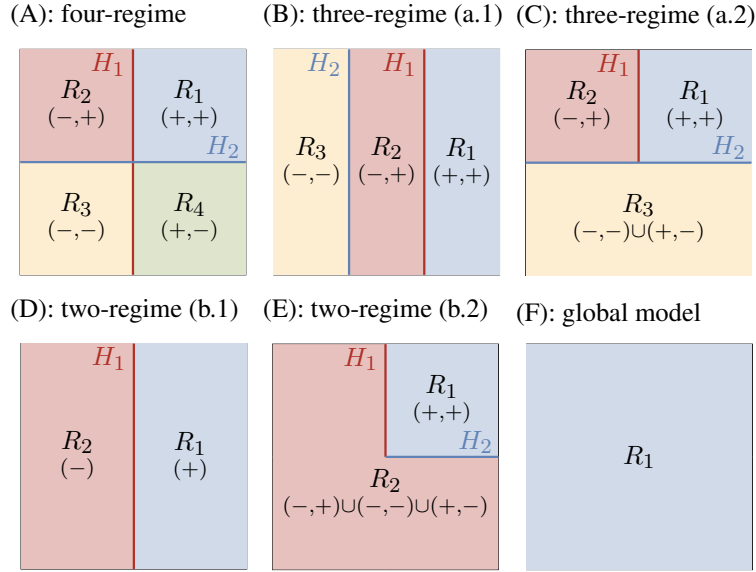

Figure \ref{fig: degenerate} illustrates the segmented models 
with no more than four regimes, which can be expressed  in a unified form 
\begin{equation}
    Y_t = \sum_{k=1}^{K_0} \bX_t\t\bbeta_{k0} \id\{ \bZ_t \in R_k(\bgamma_0)\} + \ep_t,
    \label{eq: degenerate}
\end{equation}
where the number of regimes $1 \leq K_0 < 4$ and the number of splitting hyperplanes $L_0 \leq 2$. In particular, $R_k(\bgamma_0) = \mathcal{Z}_1\times \mathcal{Z}_2$ for the global linear model ($K_0 = 1$), the splitting coefficient $\bgamma_0 = \bgamma_{10}$ or $\bgamma_{20} $ when $L_0 = 1$, and $\bgamma_0 = (\bgamma_{10}\t, \bgamma_{20}\t)\t$ when $L_0 = 2$.

Let $\what{\mathcal{B}} =\{\what{\bbeta}_k\}_{k=1}^4$ 
and $\what{\mathcal{G}} =\{\what{\bgamma}_j\}_{j=1}^2$ 
be the LS estimators for the regression and the boundary coefficients, respectively,  obtained under the four-regime regression model (\ref{eq:ls}). 
 To measure the estimation accuracy of the four-regime algorithms for less than four regime models, {we need a distance of the true parameters of {possibly} degenerated models to the set of the LS estimates under the four-regime model.  
To this end,} we define a distance between a vector 
$\bv$ and a set of vectors  
$\what{\mathcal{V}} = \{\hat{\bv}_j\}_{j=1}^J$  as $d(\bv,\what{\mathcal{V}}) = \min_j \norm{\bv-\hat{\bv}_j}_2 $.  
The following theorem establishes  the convergence of the LS estimators to the underlying parameters 
by showing that the distance of the true parameters of the degenerated models to the set of the LSEs under the four-regime model convergences to zero. 

\begin{theorem}
{For Model \eqref{eq: degenerate} with $K_0$ regimes and $L_0$ splitting hyperplanes, where $1 \leq K_0 < 4$ and $0 \leq L_0 \leq 2$, }
  under Assumption \ref{assumption: temporal} and Assumptions S2-S4 in the SM (\cite{yan-chen-2024-SM}), which adapt Assumptions \ref{assumption: for consistency}--\ref{assumption: for rate} to the degenerate model settings, {then for each $\bbeta_{k0}$ with $1 \leq k \leq K_0$,  $d(\bbeta_{k0}, \what{\mathcal{B}}) = O_p(1/\sqrt{T})$. If $L_0 = 1$, then $d(\bgamma_{0}, \what{\mathcal{G}}) = O_p(1/T)$. If  $L_0 = 2$, then $d(\bgamma_{i0}, \what{\mathcal{G}}) = O_p(1/T)$ for each $i = 1$ and $2$.}
{Moreover, for any of the degenerated models with $K_0 < 4$ regimes, there exists an 
index set $\mathcal{Q}_k \subset \{ 1, \cdots, 4\}$ 
such that $  \P\left\{\bZ \in  R_k(\bgamma_0) \bigtriangleup \cup_{i \in \mathcal{Q}_k} R_i(\what{\bgamma}) \right\} = O(1/T)$  for each $1\leq k \leq K_0$. }
      \label{thm: misspecify}
\end{theorem}

The theorem shows that under each of the degenerated models, the estimated boundaries and the regression coefficients obtained under  (\ref{eq:ls}) of the four-regime model are consistent to the true parameters in the sense of the diminishing distance between the true parameters and the sets of the estimates.  A remaining issue is to identify the true number of regimes so that more precise segmented regression can be conducted.  In the following, we introduce a model selection procedure to attain the purpose.   

{The last part of} Theorem \ref{thm: misspecify} suggests that 
each true regime $R_k(\bgamma_0)$ can either be consistently estimated by some $R_i(\what{\bgamma})$ if $|\mathcal{Q}_k| = 1$, which occurs when $R_k(\bgamma_0)$ has two boundaries, such as the first two regimes in Figure \ref{fig: degenerate} (C),  or there are some redundant estimated segments in $R_k(\bgamma_0)$, which happens if $R_k(\bgamma_0)$ has a single boundary while an unnecessary estimated hyperplane splits through $R_k(\bgamma_0)$. If the latter case is true, then $|\mathcal{Q}_k| > 1$ and  there exist two adjacent estimated regimes $R_i(\widehat{\bgamma})$ and $R_h(\widehat{\bgamma})$ with $i, h \in \mathcal{Q}_k$, whose corresponding $\what{\bbeta}_i$ and $\what{\bbeta}_h$ both consistently estimate $\bbeta_{k0}$. Under such a case, 
{merging $R_i(\widehat{\bgamma})$} with $R_h(\widehat{\bgamma})$ as one regression regime will asymptotically not lead to an increased sum of squared residuals (SSR). Otherwise, if the regression models on  $R_i(\widehat{\bgamma})$ and $R_h(\widehat{\bgamma})$ are distinct, then merging these two regimes will deteriorate the fitting performance. Such a property hints that the true model with $K_0 < 4$ can be selected via a {backward elimination} procedure.

Starting from the estimated four-regime model, we try recursively finding the best pairs of adjacent regimes to be merged, under a criterion that the merging leads to the minimal increase in the fitting errors, as defined in \eqref{eq: optimal merge} below. Via conducting the optimal regime merging recursively, we obtain four candidate regression models with the number of regimes from $K = 4$ to $K = 1$.   
In the second step, the optimal number of regimes $K$ is selected based on a criterion function \eqref{eq: select criterion} that combines a goodness-of-fit measure and a penalty for over-segmentation.

For the initial model with  four regimes, define 
\[
S_T(4) = \sum_{t=1}^T [Y_t - \sum_{k=1}^K \bX_t\t \widehat{\bbeta}_k^{ (4)} \id\{\bZ_t \in \widehat{R}_k^{(4)} \} ]^2
\]
to be the sum of square residual (SSR) of the estimated four-regime model.  
For $K = 4,3,2$, recursively define 
\bea 
&& D_T^{\scriptscriptstyle (K)}(i,h) \nn \\
&=& \min_{\bbeta \in \mathcal{B}}\sum_{t=1}^T [Y_t -  \bX_t\t {\bbeta} \id \{\bZ_t \in  \widehat{R}_i^{\scriptscriptstyle (K)}  \cup   \widehat{R}_h^{\scriptscriptstyle (K)}  \} ]^2  - \sum_{t=1}^T [Y_t - \sum_{k=i, h} \bX_t\t \widehat{\bbeta}_k^{\scriptscriptstyle (K)} \id\{\bZ_t \in \widehat{R}_k^{\scriptscriptstyle (K)}\} ]^2 \nn 
\eea 
to be  the increment in the SSR after merging $\widehat{R}_i^{\scriptscriptstyle (K)}$ and $\widehat{R}_h^{\scriptscriptstyle (K)}$. Let $\mathcal{A}_{\scriptscriptstyle K}$ be the pair of indices for the adjacent segments of $\{ \widehat{R}^{\scriptscriptstyle (K)}_k \}$. 
 We  merge
the segments $\what{R}^{\scriptscriptstyle (K)}_{\hat{i}} $ and $\what{R}^{\scriptscriptstyle (K)}_{\hat{h}}$ if 
\begin{align}
   (\what{i}, \what{h}) = \argmin_{(i,h) \in \mathcal{A}_{K}} D_T^{\scriptscriptstyle (K)}(i,h),
   \label{eq: optimal merge}
\end{align} 
followed by labeling  the merged region and the remaining regions  as $\{\what{R}^{\scriptscriptstyle (K-1)}_k \}_{k=1}^{K-1}$, and we denote
the  estimated regression coefficients to these $K-1$ regimes by $\{\what{\bbeta}_k^{\scriptscriptstyle (K-1)} \}_{k=1}^{K-1}$. Then,  define the SSR of the $(K-1)$-segment submodel as 
\[
S_T(K-1) = S_T(K) + D_T^{\scriptscriptstyle (K)} (\what{i}, \what{h}). 
\]


{After obtaining the $S_T(K)$ for $K=2,3, 4$, } we select the number of segments $\widehat{K}$ as 
\begin{align}
    \widehat{K} = \argmin_{1\leq K \leq 4} \{\log(\frac{S_T(K)}{T} ) + \frac{\lambda_T}{T}  K\}
    \label{eq: select criterion}
\end{align}
and output the estimated regimes and regression coefficients accordingly. 
The following theorem shows that the above selection algorithm has the model selection consistency. 

\begin{theorem}
   Under the assumptions of Theorem \ref{thm: misspecify}, and $\lambda_T \to \infty, \lambda_T / T \to 0$ as $T \to \infty$, then $\widehat{K}$ selected in \eqref{eq: select criterion} satisfies $\P(\widehat{K} = K_0) \to 1$ as $T \to \infty$. In addition, $\P\{\widehat{R}^{\scriptscriptstyle (\widehat{K})}_k  \bigtriangleup R_k(\bgamma_0)  \}= O(1/T)$ and $\norm{\what{\bbeta}_k^{\scriptscriptstyle (\widehat{K})} - \bbeta_{k0}} = O_p(1/\sqrt{T})$ for any $k \in \{1,\cdots, K_0\}$. 
   \label{thm: selection consistency}
\end{theorem}

Theorem \ref{thm: selection consistency} indicates that with the probability approaching $1$, the selected number of regimes $\widehat{K}$ coincides with the true number $K_0$, and as a by-product, the corresponding estimated regimes and the regression coefficients converge to their underlying counterparts.  If the regularization parameter is chosen as $\lambda_T = \log T$, the \eqref{eq: select criterion} corresponds to the Bayesian information criterion (BIC) 
\citep{schwarz1978}.

\begin{remark}     
{There are two existing approaches for carrying out the model selection for the segmented models. 
} 
 One is by conducting pairwise linearity tests. Specifically, for each adjacent regimes $R_i(\what{\bgamma})$ and $R_h(\what{\bgamma})$ under the four-regime model,  
 one can test for the hypothesis $H_0: \bbeta_{i0} = \bbeta_{h0}$  via two-regime linearity tests, such as the score-type test of \cite{yu2021}.  
 However, implementing such tests are computationally demanding, as the test statistics have to be formulated via supremum or averaging over $\bgamma \in \Gamma$, as $\bgamma$ is not identifiable under the null hypothesis of no splitting {within $R_i(\what{\bgamma})\cup R_h(\what{\bgamma})$}, which is known as the Davis problem \cite{davies1987}. 
 The other is the forward sequential fitting procedure for model selection of multi-threshold regression  models \citep{gonzalo2002estimation},  
 which requires optimization for the splitting (boundary) coefficients in each step. Compared with these two methods, the proposed model selection method has two advantages. One is that it has  quite readily computation without having to do the bootstrap for the model selection;  and the other is that 
 we only need to estimate the splitting coefficients for the initial four-segment model once and for all, as 
    the submodels with fewer regimes are selected via \eqref{eq: optimal merge} without the need to conduct non-convex optimization as in the forward sequential fitting procedure.  

\end{remark}

\section{Simulation Study}
\label{sec:simulation}

In this section, we present results {from simulation experiments designed} to investigate the performance of the proposed estimation and inference procedures for the four-regime and the degenerated less than four regime models.

\subsection{Estimation under the four-regime model}
\label{sec: four-sim}
We first conducted simulations under the four-regime model (\ref{eq:m1}) such that the sample was generated according to 
\begin{equation}
     Y_t = \sum_{k=1}^4 \bX_t\t\bbeta_{k0}
     \id_k(\bZ_{1,t}\t\bgamma_{10}, \bZ_{2,t}\t\bgamma_{20})+ \ep_t,   \quad t = 1, \cdots, T,
     \label{eq: four-sim}
\end{equation}
where $\bX_t = (\tilde{\bX}_{t}\t,1)\t$ with $\tilde{\bX}_{t}=(X_{1,t}, X_{2,t}, X_{3,t})\t$  
and $\bZ_{j,t }= (\tilde{\bZ}_{j,t}\t, 1)\t$ with $\tilde{\bZ}_{j,t} = (Z_{j,1,t}, Z_{j,2,t})\t$ 
for $j = 1,2$. {The noises were generated as $\ep_t = \sigma(\bX_t, \bZ_t) e_t$ with $\sigma(\bX_t, \bZ_t) = 1 + 0.1 X_{1,t}^2 + 0.1 Z_{1,1,t}^2$ and $\{e_t\}_{t=1}^T$ being generated independently from the standard normal distribution and independent of $\{\bX_t, \bZ_t\}_{t=1}^T$.} 
 The regression coefficients of the four regimes were $\bbeta_{10} = (1,1,1,1)\t, \bbeta_{20} = (-3, -2, -1, 0), \bbeta_{30} = (0,1,3,-1)\t$ and $\bbeta_{40} = (2, -1, 0, 2)\t$, and the two boundary coefficients $\bgamma_{10} = (1,-1,0)\t$ and $\bgamma_{20} = (1,1,0)\t$, respectively.  

  We considered three settings for $\bX_t$ and $\bZ_{j,t}$: independence, dependence with auto-regressive (AR) and moving average (MA) models, respectively.
 {Let $\bV_t = (\tilde{\bX}_t\t, \tilde{\bZ}_{1,t}\t, \tilde{\bZ}_{2,t}\t)\t$.  For the independence setting, we generated $\{\bV_t \}_{t=1}^T \overset{\mathrm{i.i.d.}}{\sim} \bm{N}(\bm{0}, \Sigma_V)$, where $\Sigma_V = (\sigma_{ij})_{i,j = 1, \cdots, 7}$ with $\sigma_{ii} = 1$ and $\sigma_{ij} = 0.1$ if $i \neq j$.
For the AR dependence, $\bV_{t} = \psi\bV_{t-1} + \bm{u}_t$, where $\{\bm{u}_t \}_{t=1}^T \overset{\mathrm{i.i.d.}}{\sim} \bm{N}(\bm{0}, \Sigma_V)$ and the dependence level $\psi \in \{ 0.2, 0.4, 0.8 \}$.  For the MA scenario, we generated $\bV_{t} = \psi\bm{u}_{t-1} + \bm{u}_t$, where $\{\bm{u}_t \}_{t=1}^T\overset{\mathrm{i.i.d.}}{\sim} \bm{N}(\bm{0}, \Sigma_V)$ and $\psi$ took values in $\{ 0.2, 0.4, 0.8 \}$, respectively.}
{The simulation experimented with four sample sizes: $\{200,400, 800, 1600\}$, and the experiments were repeated 500 times for each sample size and dependence setting. }

\begin{table}
\scriptsize  
    \caption{Empirical average estimation errors 
    $\norm{\bgamma_{0} -\what{\bgamma}}_{2}$ and $\norm{\bbeta_{0} -\what{\bbeta}}_{2}$ (multiplied by $10$),  
    under the independence (IND), auto-regressive (AR) and moving average (MA) settings {with different dependence level $\psi$} {for $\{\bX_t, \bZ_{1,t}, \bZ_{2,t}\}_{t=1}^T$}. The numbers inside the parentheses are the standard errors of the simulated averages. }
\setlength{\tabcolsep}{4pt}
{
\begin{tabular}{ccccccccccccccc}
\hline\hline
&\multicolumn{2}{c}{ IND}&\multicolumn{6}{c}{ AR}&\multicolumn{6}{c}{ MA}\\
\cmidrule(lr){2-3}  \cmidrule(lr){4-9}  \cmidrule(lr){10-15}
 \multirow{2}{*}{$T$} &\multicolumn{2}{c}{$\psi = 0$} &\multicolumn{2}{c}{$\psi = 0.2$}&\multicolumn{2}{c}{$\psi = 0.4$}&\multicolumn{2}{c}{$\psi = 0.8$} &\multicolumn{2}{c}{$\psi = 0.2$}&\multicolumn{2}{c}{$\psi = 0.4$}&\multicolumn{2}{c}{$\psi = 0.8$}\tabularnewline
\cmidrule(lr){2-3}  \cmidrule(lr){4-5} \cmidrule(lr){6-7} \cmidrule(lr){8-9} \cmidrule(lr){10-11} \cmidrule(lr){12-13} \cmidrule(lr){14-15}
&  $\what{\bgamma}$ & $\what{\bbeta}$ &  $\what{\bgamma}$ & $\what{\bbeta}$ &  $\what{\bgamma}$ & $\what{\bbeta}$ &  $\what{\bgamma}$ & $\what{\bbeta}$ &  $\what{\bgamma}$ & $\what{\bbeta}$ &  $\what{\bgamma}$ & $\what{\bbeta}$ &  $\what{\bgamma}$ & $\what{\bbeta}$ \\
\cmidrule(lr){1-3}  \cmidrule(lr){4-5} \cmidrule(lr){6-7} \cmidrule(lr){8-9} \cmidrule(lr){10-11} \cmidrule(lr){12-13} \cmidrule(lr){14-15}
  \multirow{2}{*}{$200$} 
  &0.94&6.68&0.92&6.66&0.88&6.43&0.88&5.9&0.93&6.63&0.9&6.49&0.85&6.14\tabularnewline
  & (0.59)& (1.7)& (0.58)& (1.68)& (0.6)& (1.56)& (0.61)& (2.24)& (0.56)& (1.63)& (0.54)& (1.66)& (0.52)& (1.8)\tabularnewline
\cmidrule(lr){1-3}  \cmidrule(lr){4-5} \cmidrule(lr){6-7} \cmidrule(lr){8-9} \cmidrule(lr){10-11} \cmidrule(lr){12-13} \cmidrule(lr){14-15}
  \multirow{2}{*}{$400$}  &0.45&4.55&0.45&4.55&0.45&4.4&0.43&3.98&0.44&4.46&0.43&4.38&0.43&4.06\tabularnewline
&  (0.3)& (1.1)& (0.3)& (1.11)& (0.27)& (1.17)& (0.29)& (1.53)& (0.28)& (1)& (0.33)& (1.07)& (0.28)& (1.21)\tabularnewline
\cmidrule(lr){1-3}  \cmidrule(lr){4-5} \cmidrule(lr){6-7} \cmidrule(lr){8-9} \cmidrule(lr){10-11} \cmidrule(lr){12-13} \cmidrule(lr){14-15}
  \multirow{2}{*}{$800$}  &0.25&3.11&0.24&3.09&0.22&2.97&0.22&2.64&0.23&3.11&0.25&3.03&0.22&2.81\tabularnewline
& (0.16)& (0.66)& (0.15)& (0.66)& (0.14)& (0.66)& (0.14)& (0.96)& (0.14)& (0.66)& (0.16)& (0.65)& (0.15)& (0.72)\tabularnewline
\cmidrule(lr){1-3}  \cmidrule(lr){4-5} \cmidrule(lr){6-7} \cmidrule(lr){8-9} \cmidrule(lr){10-11} \cmidrule(lr){12-13} \cmidrule(lr){14-15}
  \multirow{2}{*}{$1600$} &0.11&2.2&0.11&2.18&0.12&2.11&0.11&1.88&0.11&2.17&0.11&2.11&0.11&1.97\tabularnewline
& (0.07)& (0.46)& (0.07)& (0.47)& (0.08)& (0.5)& (0.07)& (0.77)& (0.07)& (0.45)& (0.07)& (0.47)& (0.07)& (0.54)\tabularnewline
\hline
\hline 
\end{tabular}
}
\label{tab:4-regime}
\end{table}

Table \ref{tab:4-regime} reports the average $L_2$  estimation errors under the three temporal settings (independence, AR(1) and MA(1)) {and different dependence levels $(\psi = 0.2, 0.4, 0.8)$} for $\bbeta$ 
and $\bgamma$,  
respectively. 
It {suggests that under the three dependence settings 
the estimation errors of $\what{\bgamma}$ and $\what{\bbeta}$ both decreased as the sample size $T$ was increased, indicating the convergence of the estimation in both the regression and the splitting boundary coefficients. The table also suggests that the magnitudes of the estimation errors were comparable across the three temporal settings {with different dependence levels}, {which support the result of Theorem \ref{thm: asy-dist} that the temporal dependence in $\{\bX_t, \bZ_{1,t}, \bZ_{2,t}\}_t^T$ 
does not have leading order effects on the asymptotic variance of $\what{\bbeta}$. }
Moreover,  Table \ref{tab:4-regime} shows that the simulated averages of   $\norm{\bgamma_{0} -\what{\bgamma}}_{2}$ were approximately halved once the sample size was doubled, while the reduction in 
 $\norm{\bbeta_{0} -\what{\bbeta}}_{2}$  was much slower,   confirming the faster convergence rates of $\what{\gamma}$.  
 }

\subsection{{Estimation under models with less than four regimes}}
\label{sec: sim-less}

We next investigated the performances of the proposed  estimation based on the four-regime model 
when the underlying model was degenerated with less than four regimes.  The data generating process for $\{\bX_t, \bZ_{1,t}, \bZ_{2,t}, \ep_t\}_{t=1}^T$ was largely  the independence setting used in Section \ref{sec: four-sim}. 
For the three-regime model 
\eqref{eq:3-regimes-1} with non-intersected splitting hyperplanes, we let $\bgamma_{10} = (1, 0, -1)\t, \bgamma_{20} = (1, 0,1)\t$ and $\bbeta_{10} = (1,1,1,1)\t, \bbeta_{20} = (-3,-2, -1, 0)\t, \bbeta_{30} = (0,1,3,-1)\t$. For the three-regime model   \eqref{eq:3-regimes-2} with intersected splitting hyperplanes, we let $\bgamma_{10} = (1, 1, 0)\t, \bgamma_{20} = (1, -1,0)\t$ while  $H_{10}$ does not extend to the positive side of $H_{20}$, and $\{\bbeta_{k0} \}_{k=1}^3$ were the same as above. The parameters for the two-regime model \eqref{eq:2-regimes-1} with one splitting hyperplane were set as $\bgamma_0 = (1,1,0)\t$, $\bbeta_{10} = (1,1,1,1)\t$ and $\bbeta_{20} = (-3,-2, -1, 0)\t$.  For the two-regime model \eqref{eq:2-regimes-2} with two splitting hyperplanes, we set the splitting coefficients as the same as the four-regime model \eqref{eq: four-sim}, and $R_1(\bgamma_0) = \{\bz: \bz_1\t\bgamma_{10} > 0, \bz_2\t\bgamma_{20} > 0 \}$ and $R_2(\bgamma_0) = \mathcal{Z}_1 \times \mathcal{Z}_2 \setminus R_1(\bgamma_0)$, where the regression coefficients are $\bbeta_{10} = (1,1,1,1)\t$ and $\bbeta_{20} = (-3,-2, -1, 0)\t$, respectively. Finally, the regression coefficients for the global linear model \eqref{eq:1-regime} were $\bbeta_0 = (1, 1, 1, 1)\t$.

The simulation results {are} reported in Tables S2 of Section H.2 in the SM (\cite{yan-chen-2024-SM}). 
They show that for all the models with less than four regimes,  the empirical averages of $\sum_i d(\bgamma_{i0}, \what{\mathcal{G}})$ and $\sum_{k} d(\bbeta_{k0}, \what{\mathcal{B}})$ all diminished to $0$ at similar rates as those in Table \ref{tab:4-regime}, where $\what{\mathcal{G}}$ and $\what{\mathcal{B}}$ are the sets of estimators obtained under the four-regime model for the splitting and regression coefficients, respectively. 
These confirmed the results in Theorem \ref{thm: misspecify}. 
 In addition, to evaluate the cost of not knowing the number of the underlying regimes, we also estimated $\bgamma_0$ and $\bbeta_0$ in the oracle setting, in which the true model forms were known. It was found that estimation errors of $\bgamma_0$ under the four-regime model fitting were about the same as that obtained under the oracle models, which was because the four-regime estimator can efficiently use the data points located near the underlying boundaries as the oracle estimators did. Moreover, as shown in Figures S2 and S3 of the SM (\cite{yan-chen-2024-SM}), if the estimated four-regime model produced redundant segments within a true regime, then the discrepancy between the estimated regression coefficients on these redundant segments converged to $0$, which verified the idea used in the optimal merger strategy for the backward elimination procedure in the model selection. 

\subsection{Model selection}
\label{sec: sim selection}

We then conducted simulation experiments to examine the performance of the proposed model selection method in Section \ref{sec: Model Specification}.  
We considered the true number of regimes ranging from $K_0 = 4$ to $K_0 = 1$,  
where the  parameters for the model with $K_0 = 4$ were the same as Model (7.1) 
and those for $K_0 = 3$ and $K_0 = 2$ were Model \ref{eq:3-regimes-1} and Model \ref{eq:2-regimes-1}, respectively, in Section \ref{sec: sim-less}. 
More simulation results for Model \eqref{eq:3-regimes-2} and Model \eqref{eq:1-regime} ($K_0 = 1$) were reported in Table S3 of the SM.

\begin{table}[ht]
\scriptsize  
\caption{Empirical model selection results under $500$ replications. The performances were evaluated by the average estimated number of regimes $\what{K}$, the discrepancy between the true regimes and the estimated regimes  $D(\mathcal{R}, \what{\mathcal{R}})$ and the  $L_2$ estimation error of regression coefficients $D(\mathcal{B}, \what{\mathcal{B}})$. The penalty parameter $\lambda_T$ 
was chosen in $ \{ 5, 5 \log(T),5\log^2(T)\}$.  The numbers inside the parentheses are the standard errors of the simulated averages.  }
\begin{center}
\begin{tabular}{c c ccc ccc ccc}
\hline\hline
\multirow{2}{*}{Model} & \multirow{2}{*}{$T$}  &\multicolumn{3}{c}{$\lambda_T = 5$} &\multicolumn{3}{c}{$\lambda_T = 5\log(T)$} &\multicolumn{3}{c}{$\lambda_T = 5\log^2(T)$} \\
 \cmidrule(lr){3-5} \cmidrule(lr){6-8} \cmidrule(lr){9-11} 
& &  $\widehat{K}$ & $D(\mathcal{R}, \what{\mathcal{R}}) $& $D(\mathcal{B}, \what{\mathcal{B}})$ &  $\widehat{K}$ & $D(\mathcal{R}, \what{\mathcal{R}}) $& $D(\mathcal{B}, \what{\mathcal{B}})$  &  $\widehat{K}$& $D(\mathcal{R}, \what{\mathcal{R}}) $& $D(\mathcal{B}, \what{\mathcal{B}})$   \\ 
\hline
 \multirow{8}{*}{\makecell{Model \eqref{eq:m1}  \\ ($K_0 = 4$)}}& \multirow{2}{*}{200 } 
 &4.00&0.03&0.61&3.99&0.03&0.62&2.78&0.87&2.24\tabularnewline
& &(0.00)&(0.02)&(0.12)&(0.08)&(0.04)&(0.16)&(0.87)&(0.91)&(1.05)\tabularnewline
 \cmidrule(lr){2-5} \cmidrule(lr){6-8} \cmidrule(lr){9-11} 
&\multirow{2}{*}{400 } &4.00&0.01&0.41&4.00&0.01&0.41&3.92&0.05&0.53\tabularnewline
& &(0.00)&(0.01)&(0.08)&(0.00)&(0.01)&(0.08)&(0.27)&(0.13)&(0.43)\tabularnewline
 \cmidrule(lr){2-5} \cmidrule(lr){6-8} \cmidrule(lr){9-11} 
&\multirow{2}{*}{800 } &4.00&0.01&0.29&4.00&0.01&0.29&4.00&0.01&0.29\tabularnewline
& &(0.00)&(0.00)&(0.05)&(0.00)&(0.00)&(0.05)&(0.00)&(0.00)&(0.05)\tabularnewline
 \cmidrule(lr){2-5} \cmidrule(lr){6-8} \cmidrule(lr){9-11} 
&\multirow{2}{*}{1600 }&4.00&0.00&0.20&4.00&0.00&0.20&4.00&0.00&0.20\tabularnewline
& &(0.00)&(0.00)&(0.04)&(0.00)&(0.00)&(0.04)&(0.00)&(0.00)&(0.04)\tabularnewline
\hline 
 \multirow{8}{*}{\makecell{Model \eqref{eq:3-regimes-1}  \\ ($K_0 = 3$)}}& \multirow{2}{*}{200 } 
& 3.44&0.12&0.50&3.00&0.02&0.48&2.85&0.13&0.75\tabularnewline
 & &(0.50)&(0.11)&(0.11)&(0.00)&(0.02)&(0.11)&(0.38)&(0.30)&(0.69)\tabularnewline
  \cmidrule(lr){2-5} \cmidrule(lr){6-8} \cmidrule(lr){9-11} 
 &\multirow{2}{*}{400 }  &3.39&0.10&0.34&3.00&0.01&0.33&3.00&0.01&0.33\tabularnewline
 & &(0.49)&(0.11)&(0.07)&(0.00)&(0.01)&(0.07)&(0.00)&(0.01)&(0.07)\tabularnewline
  \cmidrule(lr){2-5} \cmidrule(lr){6-8} \cmidrule(lr){9-11} 
 &\multirow{2}{*}{800 } &3.33&0.08&0.23&3.00&0.01&0.22&3.00&0.01&0.22\tabularnewline
 & &(0.47)&(0.11)&(0.05)&(0.00)&(0.00)&(0.05)&(0.00)&(0.00)&(0.05)\tabularnewline
  \cmidrule(lr){2-5} \cmidrule(lr){6-8} \cmidrule(lr){9-11} 
 &\multirow{2}{*}{1600 } &3.33&0.08&0.16&3.00&0.00&0.16&3.00&0.00&0.16\tabularnewline
& &(0.47)&(0.11)&(0.03)&(0.00)&(0.00)&(0.03)&(0.00)&(0.00)&(0.03)\tabularnewline
\hline
\multirow{8}{*}{\makecell{Model \eqref{eq:2-regimes-1}  \\ ($K_0 = 2$)}}& \multirow{2}{*}{200 } &3.38&0.14&0.35&2.03&0.01&0.30&2.00&0.01&0.30\tabularnewline
&&(0.59)&(0.11)&(0.10)&(0.17)&(0.01)&(0.08)&(0.00)&(0.01)&(0.08)\tabularnewline
   \cmidrule(lr){2-5} \cmidrule(lr){6-8} \cmidrule(lr){9-11} 
&\multirow{2}{*}{400 } &3.54&0.13&0.24&2.01&0.01&0.20&2.00&0.01&0.20\tabularnewline
&&(0.51)&(0.11)&(0.07)&(0.08)&(0.01)&(0.05)&(0.00)&(0.00)&(0.05)\tabularnewline
   \cmidrule(lr){2-5} \cmidrule(lr){6-8} \cmidrule(lr){9-11} 
&\multirow{2}{*}{800 } &3.53&0.12&0.16&2.00&0.00&0.14&2.00&0.00&0.14\tabularnewline
&&(0.53)&(0.11)&(0.04)&(0.06)&(0.00)&(0.04)&(0.00)&(0.00)&(0.04)\tabularnewline
   \cmidrule(lr){2-5} \cmidrule(lr){6-8} \cmidrule(lr){9-11} 
&\multirow{2}{*}{1600 } &3.50&0.13&0.12&2.00&0.00&0.10&2.00&0.00&0.10\tabularnewline
&&(0.55)&(0.12)&(0.03)&(0.00)&(0.00)&(0.03)&(0.00)&(0.00)&(0.03)\tabularnewline
\hline
\hline
\end{tabular}\end{center}
\label{tab: selection}
\end{table}

Table \ref{tab: selection} reports three model selection performance measures for the simulation, 
namely 
(i) 
the estimated number of regimes $\widehat{K}$,  (ii) the discrepancy between the true regimes and the estimated regimes measured by 
$$
D(\mathcal{R}, \what{\mathcal{R}}) =   \sum_{k=1}^{K_0} \min_{1\leq h \leq \what{K}} \left\{T^{-1} \sum_{t=1}^T \abs{\id\{\bZ_t \in R_k(\bgamma_0)\} -\id\{\bZ_t \in R_h(\what{\bgamma})\}} \right\},
$$ 
where $\mathcal{R} = \{R_k(\bgamma)\}_{k=1}^{K_0}$ and $\what{\mathcal{R}} = \{R_k(\what{\bgamma})\}_{k=1}^{\hat{K}}$,  and  (iii) the $L_2$ estimation error of regression coefficients, quantified by $D(\mathcal{B}, \what{\mathcal{B}}) = \sum_{k=1}^{K_0}  \min_{1\leq h \leq \what{K}} \norm{\bbeta_{k0} - \what{\bbeta}_h}$. To evaluate the impact of the penalty parameter $\lambda_T$ in \eqref{eq: select criterion}, we presented the results under three different choices:  $\lambda_T = 5, 5 \log(T)$ and $5\log^2(T)$.  

Table \ref{tab: selection} shows that, for the constant penalty $\lambda_T = 5$, although the estimated number of regimes $\what{K}$ was consistent under $K_0 = 4$,  it tended to select overly segmented models when $K_0 < 4$. Both $\lambda_T = 5 \log(T)$ and $5\log^2(T)$ led to consistent estimated $\what{K}$ for all models, which confirmed the assertion in Theorem \ref{thm: selection consistency} that $\lambda_T$ satisfying $\lambda_T \to \infty$ and $\lambda_T /T \to 0$ leads to model selection consistency. It was also noted that while the last two penalties were consistent, for smaller sample sizes, the selection performance with $\lambda_T = 5 \log (T)$ was superior to that with $\lambda_T = 5 \log^2(T)$ when $K_0 \geq 3$, while the latter penalty had better selection accuracy when $K \leq 2$. Such a phenomenon may be understood since a larger penalty tends to encourage under-segmentations. In addition, both  $D(\mathcal{R}, \what{\mathcal{R}}) $ and $D(\mathcal{B}, \what{\mathcal{B}})$ diminished to $0$ when $\widehat{K}$ was correctly selected, indicating that the model specification procedure was able to not only consistently identify $K_0$, but also led to consistent estimates of regimes and the corresponding regression coefficients, as shown in Theorem \ref{thm: selection consistency}.

\subsection{Smoothed regression bootstrap}

We now report simulation results designed to evaluate the empirical performance of the smoothed regression bootstrap.  

{The data generating model for $\{Y_t, \bX_t, \bZ_{1,t}, \bZ_{2,t}\}_{t=1}^T$ was the same as the independent setting in Section \ref{sec: four-sim}, but $(\tilde{\bX}_t\t, \tilde{\bZ}_{1,t}\t, \tilde{\bZ}_{2,t}\t)\t$ was truncated over a 7-dimensional region $[-2, 2]^7$ to ensure the distribution of the covariates was compactly supported as required in Assumption \ref{assumption: smooth bootstrap}.}
The product Gaussian kernel was used as the kernel function 
{with the smoothing bandwidths $h_i$ and $b_i (i = 1, 2)$ for $\tilde{F}_0(\bx, \bz)$ and $\tilde{\sigma}^2(\bx, \bz)$  were chosen by the cross-validation method (\cite{fan-yao-1998}).  } 
As a comparison, we also conducted the wild bootstrap procedure (\cite{liu1988}), which is a commonly used bootstrap method in regression. Different from the smoothed regression bootstrap, the wild bootstrap does not resample the covariates  
and the resampled residuals  $\ep_t^* = d_t^* \what{\ep}_t$, where $\what{\ep}_t$ was the estimated residual and $d_t^*$ followed a two-point distribution. 
Both the smoothed regression bootstrap and the wild bootstrap were based on  $B = 500$ resamples for each simulation run. As there are infinitely many solutions for $\hat{\bgamma}$ from the MIQP algorithm,  for each bootstrap resample, we outputted $N=100$ solutions for the LSE of $\bgamma_0$ and used their average as $\what{\bgamma}^{*c}_b$. 

\begin{table}[ht]
\caption{Empirical coverage probabilities and widths ($\times 100$ in parentheses) of the $95\%$ confidence intervals for five projected parameters $\{\tilde{\bgamma}\t\bm{d}_i\}_{i=1}^5$ obtained with the smoothed regression bootstrap (Smooth) and the wild bootstrap (Wild) based on $500$  resamples.}
    \centering
        \begin{tabular}{c cc cc cc cc cc}
    \hline \hline 
        \multirow{2}{*}{$T$}  & \multicolumn{2}{c}{$\bm{d}_1$} & \multicolumn{2}{c}{$\bm{d}_2$} & \multicolumn{2}{c}{$\bm{d}_3$}& \multicolumn{2}{c}{$\bm{d}_4$} & \multicolumn{2}{c}{$\bm{d}_5$}  \\
         \cmidrule(lr){2-3} \cmidrule(lr){4-5} \cmidrule(lr){6-7} \cmidrule(lr){8-9}\cmidrule(lr){10-11}
         & Smooth & Wild  & Smooth & Wild & Smooth & Wild & Smooth & Wild& Smooth & Wild \\ 
  \cmidrule(lr){1-3} \cmidrule(lr){4-5} \cmidrule(lr){6-7} \cmidrule(lr){8-9}\cmidrule(lr){10-11}        
  \multirow{2}{*}{200 } & 0.92 & 0.87 & 0.97 & 0.87 & 0.93 & 0.90 & 0.93  & 0.83 & 0.96  & 0.86\\ 
         & (6.76) & (3.57) & (6.91) & (3.91) & (5.78) & (4.02) & (6.20) & (3.44)  & (6.86) & (3.56)\\ 
           \cmidrule(lr){1-3} \cmidrule(lr){4-5} \cmidrule(lr){6-7} \cmidrule(lr){8-9}\cmidrule(lr){10-11}        
          \multirow{2}{*}{400 } & 0.95 & 0.86 &  0.94 & 0.83 & 0.97 & 0.86 & 0.94& 0.88 & 0.97 &  0.85\\ 
         & (3.31) & (1.69) & (3.57) & (1.89) & (2.56) & (1.94) & (3.37) & (1.73)& (3.69) & (1.75) \\ 
           \cmidrule(lr){1-3} \cmidrule(lr){4-5} \cmidrule(lr){6-7} \cmidrule(lr){8-9}\cmidrule(lr){10-11}        
        \multirow{2}{*}{800 } & 0.93 & 0.85 & 0.96 & 0.87 & 0.94 & 0.88 & 0.96 & 0.88 & 0.96 & 0.87 \\ 
         & (1.70) & (0.83) & (1.76) & (0.99) & (1.68) & (1.00) & (1.72) & (0.86)& (1.80) & (0.76) \\ 
           \cmidrule(lr){1-3} \cmidrule(lr){4-5} \cmidrule(lr){6-7} \cmidrule(lr){8-9}\cmidrule(lr){10-11}        
           \multirow{2}{*}{1600 } &0.95  & 0.83  & 0.94& 0.88  & 0.95 & 0.90 & 0.96 &0.84  & 0.94 & 0.85 \\ 
         & (0.81) & (0.40) & (0.86) & (0.51) & (0.89) & (0.53) & (0.85) & (0.41)& (0.79) & (0.42) \\ 
         \hline \hline
    \end{tabular}
    \label{tab: bootstrap}
\end{table}

To evaluate the quality of the two bootstrap schemes, we constructed $95\%$ confidence intervals (CIs) 
for $\tilde{\bgamma}_0 = (\bgamma_{-1, 10}', \bgamma_{-1,20}')\t= (-1,0, 1, 0)\t$ projected on {five} directions $\{\bm{d_i}\}_{i=1}^5$ 
where $\bm{d_i} = \bm{e}_{i}$ for $i=1,\dots, 4$ and $\bm{d}_5 = \sum_{i = 1}^4 \bm{d}_i / 2$, and $\bm{e}_i = (e_{i1}, \cdots, e_{i4})\t$ with $e_{ii} = 1$ and $e_{ij} = 0$ if $j \neq i$.  
Table \ref{tab: bootstrap} reports the coverage probabilities and widths of the nominal $95\%$ CIs for $\tilde{\bgamma}_0\t \bm{d}_i$ based on the smoothed regression bootstrap and the wild bootstrap, respectively. It is shown that the smoothed regression bootstrap had satisfactory coverage as its empirical coverage levels were quite close to the nominal $95\%$ level under large sample sizes for all the five projection directions. This verified the consistency of the proposed bootstrap procedure in Theorem \ref{thm: smooth bootstrap}. 
On the other hand, the wild bootstrap had substantial under-coverage, 
and its coverage was not improved with the increases of the sample sizes. The comparison between the two bootstrap schemes reveals that for the inference of $\bgamma_0$, it is crucial to conduct resampling from a smoothed distribution, 
as advocated in Section \ref{sec: gamma inference}.
\section{Case Study}
\label{sec:case study}
{Air quality} is naturally affected by meteorological regimes as the latter defines the atmospheric dispersion conditions.  
We demonstrate here that the four-regime regression model is {well suited for  PM$_{2.5}$  modeling in Beijing.} 

We considered hourly  PM$_{2.5}$ data from Wanshouxigong site in central Beijing with 
the  {meteorological data from the nearest weather observation site being used. The study period was from December 1, 2018 to November 30, 
2019, which encompassed four seasons.  
 The meteorological data included the air temperature (TEMP), dew point temperature (DEWP), surface air pressure (PRES), the cumulative wind speed (IWS) at a direction and wind direction (WD).  
Cumulative rainfall (RAIN) 
was  included 
in summer, however not in the other three seasons due to a lack of it.  The categorical wind direction (WD)  took five values: Northwesterly (NW), Northeasterly (NE), Southwesterly (SW), Southeasterly (SE) and calm and variable (CV). 
We also used the boundary layer height (BLH), which defines the vertical dispersion property,  
from European Centre for Medium-Range Weather Forecasts (ECMWF).  

To investigate the in-sample and out-of-sample performances, 
the data {were} divided to the training and testing sets,  where the testing sets consisted of the data from the 11-th to the 20-th days of a month and the training sets included the rest of the data in the month.  
PM$_{2.5}$ was regressed on 
covariates  TEMP, DEWP, PRES, $\log(\text{BLH})$, IWS,
WD as well as the  PM$_{2.5}$ at the previous hour {(Lag PM$_{2.5}$)}. 
{For the wind direction,  NW, NE, SW and SE were set as dummy covariates with the CV as the baseline. 

Along with the proposed four-regime model (4-REG),  the global linear regression (GLR), {the} two-regime  model (2-REG)  \citep{lee2021} and \cite{yu2021}, the linear regression tree (LRT) 
(\cite{ zeileis2008}) and the multivariate adaptive regression splines (MARS)  \citep{friedman1991} were also considered. 
For 2-REG and 4-REG, the splitting boundaries were determined by  
TEMP, DEWP, $\log(\text{BLH})$, IWS, and the four wind directions NE, NW, SE and SW  
with the coefficients {standardized so that the intercept term being $1$.} 

 \begin{figure}[ht]
    \centering
    \caption{\small Mean squared errors (MSE) for PM$_{2.5}$ on the training (red) and testing (green) sets for each season of five models, including global linear regression (GLR),  two-regime model (2-REG), four-regime model (4-REG), linear regression tree (LRT) and multivariate adaptive regression splines (MARS), with model ranks (in increasing order of the MSEs) marked on top of the bars.  }
    \includegraphics[width=0.9\textwidth, height = 13em]{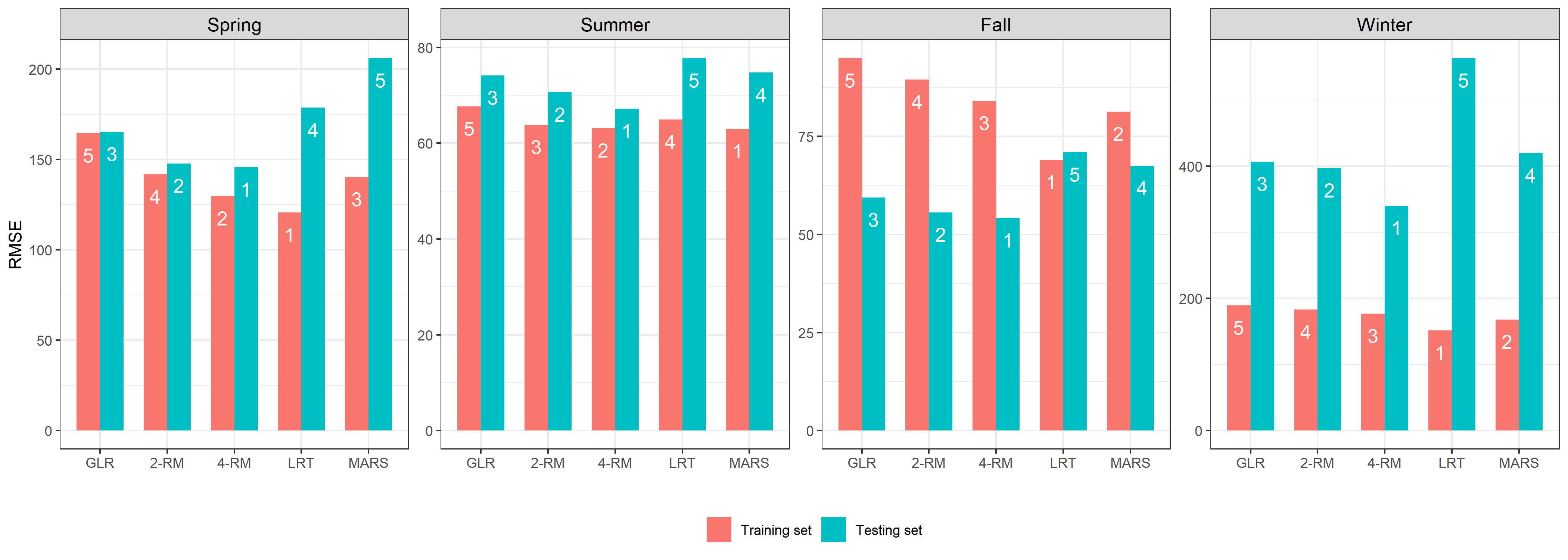}
    \label{fig:mse}
\end{figure}

Figure \ref{fig:mse} summarizes the in-sample and out-of-sample MSEs of these  models in each season. Within the training sets, LRT or MARS  achieved the lowest MSE 
among the five {models} with the average rank being 1.75 and 2, respectively. Here, rank 1 indicates the best performance. 
The average rank of the 4-REG in the training groups was 2.5, while those of the 2-REG and GLR ranked the lowest in all seasons.  
However, LRT and MARS had the highest prediction MSEs on the testing sets, even worse than the benchmark GLR for all seasons, indicating they were severely 
 over-fitted.  The segmented linear models,  
 4-REG and 2-REG,  were the best two in terms of out-of-sample performances, with the 4-REG achieving the lowest predictive errors consistently in all seasons. 

The estimated  4-regimes models in the spring, summer and fall seasons all had three regimes, 
as the fourth estimated regime had zero sample size in the three seasons.  
{A further examination suggested} 
that the two estimated boundaries had no intersections over the sample regions, which corresponded to Model (\ref{eq:3-regimes-1}) and reflected the fact that the proposed LS criterion based on the four-regime model may be able to produce a three-regime model if the latter offers better fit.  The winter had four estimated regimes.  
The estimated regression coefficients and their $95\%$ confidence intervals 
are given in Figure S4 of the SM (\cite{yan-chen-2024-SM}).

\begin{table}[ht]
 \caption{Estimated coefficients of the splitting boundaries and $\cos$ of the angle $\phi$ between the two boundaries. The coefficients were normalized such that the coefficients of the intercept terms were $1$. All the covariates were standardized such that their sample means were $0$ and standard deviations were $1$ in each season. }
  \centering 
  \bgroup
\begin{tabular}{l|c| cccccccc |c }
  \hline\hline
Season & $\bgamma$  & TEMP & DEWP & IWS & $\log(\text{BLH})$ & NE & NW & SE & SW & $\cos{\phi}$\\ 
  \hline
\multirow{2}{*}{Spring} & 1 & 1.3 & -2.5 & -0.0 & -0.4 & 0.9 & 0.3 & 0.1 & 0.0  & \multirow{2}{*}{0.78} \\ 
 & 2 & 0.4 & -0.5 & -0.1 & -0.1 & 0.6 & 0.6 & 0.1 & 0.3 &  \\ 
 \hline
  \multirow{2}{*}{Summer} & 1  &1.0 & 5.5 & -12.9 & -0.0 & -12.7 & -15.0 & -8.9 & -9.0  &  \multirow{2}{*}{0.75} \\ 
   & 2  &0.4 & 0.2 & -0.2 & 0.0 & -0.7 & -0.7 & -0.7 & -0.7  &  \\ 
   \hline 
   \multirow{2}{*}{Fall} & 1  & 0.7 & -1.0 & 0.3 & -0.1 & 0.5 & -0.0 & 0.3 & 0.0  &\multirow{2}{*}{0.65} \\ 
  & 2  & -0.5 & 1.6 & -1.0 & 0.0 & 0.1 & -1.6 & -1.3 & -0.1& \\ 
  \hline 
  \multirow{2}{*}{Winter} & 1  & 0.2 & -0.5 & 0.6 & -0.2 & 0.2 & 0.4 & 0.4 & -0.4 & \multirow{2}{*}{0.45}\\ 
   & 2  & 0.0 & -0.6 & 0.2 & -0.4 & 1.2 & 1.4 & 0.3 & 1.0  & \\ 
   \hline\hline
\end{tabular}
  \egroup
\label{tab:gamma coef}
\end{table}

Table \ref{tab:gamma coef}  reports the estimated coefficients of the two splitting boundaries for each season as well as the cosine of the dihedral  angle (denoted as $\phi$) 
between the two boundary {hyperplanes}. 
It can be seen that $\cos \phi$ for the first three seasons were relatively larger than {that in}  winter, which explains why the boundary hyperplanes of these three seasons were non-intersected. 
Table \ref{tab:gamma coef} indicates that the DEWP and the wind-related variables were the most 
influential in determining the slopes of the estimated boundaries due to their absolute coefficient values as the $\bgamma$ was normalized.  This 
reveals an attraction of the proposed regime-splitting mechanism in that the splitting boundaries are determined empirically by multivariate covariates, which contrasts to the  threshold regression where the boundary variable has to be user-specified. 
 \begin{figure}[h!]
 \caption{\small {Bar and rose plots for key variables   under each  
 estimated regimes in spring and fall 2019.}
 The height of the bars indicate the sample means with imposed line segments indicating twice of the sample deviations above and below the means. 
 The rose plots display the distribution of wind directions (width of angles) and average speed (length of
radius). Sample sizes of each regime is reported in the subtitle.}
\hspace*{-3cm}            
\begin{subfigure}{\textwidth}
   \centering
   \caption{\centering Spring}
   \includegraphics[height =0.23\textheight]{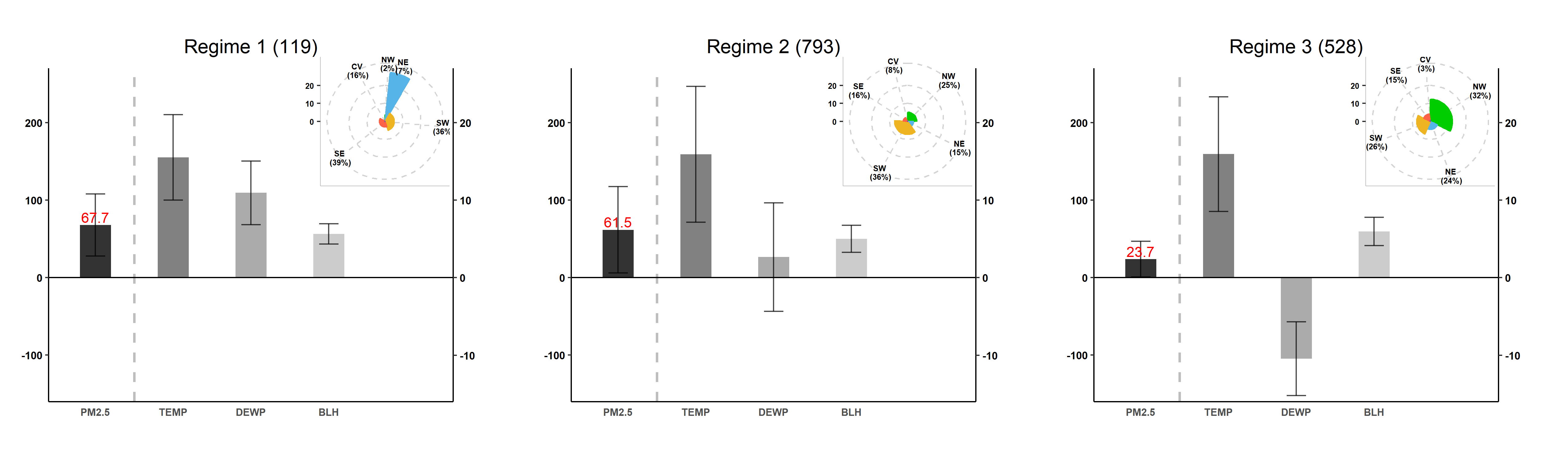}
 \end{subfigure}
 \hspace*{-3cm}            
  \begin{subfigure}{\textwidth}
   \centering
     \caption{\centering Fall}
   \includegraphics[height =0.23\textheight]{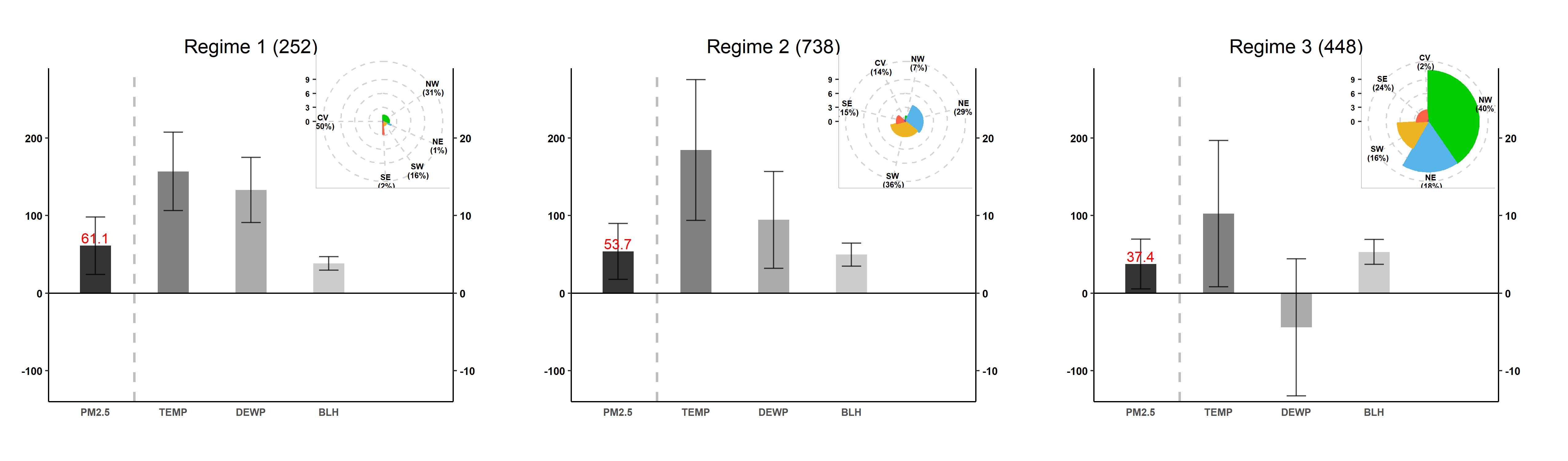}
 \end{subfigure}
\label{fig:regime_dist}
\end{figure} 

Figure \ref{fig:regime_dist} displays summary statistics of PM$_{2.5}$ and the meteorological variables under the three regimes in the spring  and fall seasons,  
as well as the rose plots for 
the wind directions and the average integrated wind speed (IWS). 
It shows that the segmented regression picked up three meteorological regimes 
on PM$_{2.5}$ where 
{Regime 1 corresponded to the pollution state with high DEWP and high proportion of Calm and Variable wind (CV) which are known to encourage the secondary
generation of PM2.5 and unfavorable static atmospheric diffusion, Regime 2 was a transitional state between the clean and high pollution states with reduced DEWP and CV, and Regime 3 was a cleaning state dominated by the northerly wind which brought cleaner and cooler air from the north. Results of the other {two} seasons and analysis are provided in Figure S5 of the SM. 
}
\section{Discussion}
\label{sec: discussion}
This paper develops a statistical inference approach for four-regimes segmented linear models, which broadens the scope of the two-regime models of \cite{lee2021} and \cite{yu2021}, and can attains valid inference for degenerated models with less than four regimes.  The proposed segmented model  is   shown to produce better in-sample and out-sample results for the air quality data in Beijing and  produced regime-splitting results which had clear atmospheric physics interpretation.  

There are two possible extensions which may be considered in future research. 
One is to allow endogeneity which may be encountered in economic and social behavior applications. 
If $\bX_t$ is endogenous and $\bZ_t$ is exogenous, $\bbeta_0$ and $\bgamma_0$ can be consistently estimated with instrument variables $\bV_t$ and the two-stage least squares estimation (2SLS) {by first regressing $\bX_t$ on $\bV_t$, and then using the fitted $\hat{\bX}_t$ to substitute $\bX_t$ in the four-regime model.  
The LS estimation via the MIQP and the inference methods for the four-regime model presented in this paper is still applicable.}
However, the 2SLS is no longer working  if   $\bZ_t$ is endogenous as discussed in \cite{yu-phillips-2018}, who proposed 
a conditioning and re-centering approach  
which might be extended to the four-regime model. 
Specifically, let $g(\bX_t, \bZ_t) =   \bX_t\t \bbeta_{10} + \E(\ep_t|\bX_t, \bZ_{t})$, $\bdelta_{k0} = \bbeta_{k0} - \bbeta_{10}$ for $k \neq 1$,  and $e_t = \ep_t -\E(\ep_t|\bX_t, \bZ_{t}) $, then Model \eqref{eq:m1} can be written as  
$ Y_t = g(\bX_t, \bZ_t) +  \sum_{k=1}^{3} \bX_t\t\bdelta_{k0} \id\{\bZ_t \in R_k(\bgamma_0) \} + e_t$,   
which is a partially linear segmented model, where $\bgamma_0$ is  
identifiable without instrument variables. 
{However, the integrated difference kernel estimator used in \cite{yu-phillips-2018} was designed for univariate threshold, and it is interesting to see how it can be extended to multivariate $\bgamma_0$.  Alternatively, one may consider estimating $\bgamma_0$ via the mixed integer programming with the nonlinear $\E(\ep_t|\bX_t, \bZ_{t})$ part approximated via sieve functions. How to solve these issues in the context of the four-regime model requires further investigation.}
}

{
Another extension is 
for segmented models with $L > 2$ splitting hyperplanes. 
{In general, the $L$ splitting hyperplanes in $\R^d$ can lead to {as many as} $K_{L} = \sum_{i=0}^{\min(L,d)} \binom{L}{i}$ segments, as shown in Section G of the SM (\cite{yan-chen-2024-SM}). 
{
It is clear that the investigations in this study for the two boundary case provide vital understanding to the general cases.
{For example, if we consider an extension to the case of having three hyperplanes in $\R^d$, we can fit a segmented model with $K = \sum_{i=0}^{\min(3,d)} \binom{3}{i}$ regimes by the least squares estimation, whose criterion function would have the same form as \eqref{eq:ls-1}. The backward selection procedure in Section \ref{sec: Model Specification} can be employed to specify the optimal number of regimes, and the smoothed regression bootstrap is still able to facilitate the inference for $\bgamma_0$ and $\bbeta_0$. Furthermore, the proof for the asymptotic distributions of the least squares estimators can be modified to suit the more general segmented models. The main challenge for the general cases is the complicated model form and demanding computation costs caused by the increase of $L$, requiring efforts in further studies.}
On the other hand, as $K_L$ grows exponentially with respect to $L$ if $d > L$ and polynomially if $d \leq L$,  there would be little need to consider segmented models with large $L$ and $d$ 
as the nonparametric local models (regression trees, etc) may be better suited. 


%
}}

\section*{Acknowledgements}
The research was partially supported by National Natural Science Foundation of China grants  12292980, 12292983 {and 92358303}.  

\begin{supplement}
\stitle{Supplement to ``Statistical Inference on  Four-Regime Segmented Regression Models''}
\sdescription{In the supplementary material, we present technical details, proofs and additional results of the simulations and the case study.
}
\end{supplement}



\bibliographystyle{imsart-number}  
\bibliography{reference}       

\end{document}